\documentclass[fleqn,usenatbib,useAMS]{mnras}

\usepackage{amsmath}
\usepackage{color}
\usepackage{subfigure}
\usepackage{float}

\RequirePackage{fixltx2e}

\def\la{\mathrel{\hbox{\rlap{\hbox{\lower4pt\hbox{$\sim$}}}\hbox{$<$}}}}

\def\ga{\mathrel{\hbox{\rlap{\hbox{\lower4pt\hbox{$\sim$}}}\hbox{$>$}}}}

\usepackage{journal_names}
\usepackage{color}

\usepackage{url}

\newcommand\farcss{\mbox{$.\!\!\!^{\prime\prime}$}}
\def\farcm{\mbox{.\kern -0.5ex\raisebox{.6ex}{\scriptsize$\prime$}}}

\def\farcss{
 \mbox{ 
  \kern  0.13ex. 
   \kern -0.95ex\raisebox{.6ex}{\scriptsize$\prime\prime$}
  \kern -0.1ex
 }
}

\usepackage{graphicx,times}

\usepackage{float}

\title[Binary stars: determining the formation location of planets]{Stellar abundance of binary stars: their role in determining the formation location of super-Earths and ice giants}
\author[~Bitsch et al.]
{\parbox{\textwidth}{B.~Bitsch$^{1}$\thanks{E-mail: bitsch@mpia.de},
R.~Forsberg$^{2}$,~ 
F.~Liu$^{2}$,~
A.~Johansen$^2$~} \vspace{0.4cm}\\
\parbox{\textwidth}{$^{1}$Max-Planck-Institut f\"ur Astronomie, K\"onigstuhl 17, 69117 Heidelberg, Germany\\
$^{2}$Lund Observatory, Department of Astronomy and Theoretical Physics, Lund University, 22100 Lund, Sweden} }

\begin{document}

\pagerange{\pageref{firstpage}--\pageref{lastpage}} \pubyear{2018}

\maketitle

\label{firstpage}

\begin{abstract}
Binary stars form from the same parent molecular cloud and thus have the same chemical composition. Forming planets take building material (solids) away from the surrounding protoplanetary disc. Assuming that the disc's accretion onto the star is the main process that clears the disc, the atmosphere of the star will show abundance reductions caused by the material accreted by the forming planet(s). If planets are only forming around one star of a binary system, the planet formation process can result in abundance differences in wide binary stars, if their natal protoplanetary discs do not interact during planet formation. Abundance differences in the atmospheres of wide binaries hosting giant planets have already been observed and linked to the formation location of giant planets. Here, we model how much building material is taken away for super-Earth planets that form inside/outside of the water ice line as well as ice giants forming inside/outside of the CO ice line. Our model predicts a significant abundance difference $\Delta$[X/H] in the stellar atmospheres of the planet-hosting binary component. Our model predicts that super-Earths that form inside the water ice line ($r<r_{\rm H_2O}$) will result in an $\Delta$[Fe/H]/$\Delta$[O/H] abundance difference in the their host star that is a factor of 2 larger  than for super-Earths formed outside the water ice line ($r>r_{\rm H_2O}$) in the water rich parts of the disc. Additionally, our model shows that the $\Delta$[Fe/H]/$\Delta$[C/H] abundance difference in the host star is at least a factor of 3 larger for ice giants formed at $r<r_{\rm CO}$ compared to ice giants formed far out in the protoplanetary disc ($r>r_{\rm CO}$). Future observations of wide binary star systems hosting super-Earths and ice giants could therefore help to constrain the migration pathway of these planets and thus constrain planet formation theories.
\end{abstract}


\section{Introduction}
\label{sec:Introduction}

After the first observation of an exoplanet more than 20 years ago \citep{1995Natur.378..355M}, new methods and observations have revealed several thousands of exoplanets. Within these found planets are two groups of planets that are not harboured in the solar system: hot Jupiters and super-Earths. Jupiter type planets, however, are very rare, where only about $\sim$ 10\% of stars harbour gas giants \citep{2010PASP..122..905J, 2011arXiv1109.2497M}, where the occurrence rate of wide orbit gas giants is much higher than for close in gas giants \citep{2016ApJ...819...28W, 2016AJ....152..206F}. In addition, the occurrence rate of giant planets increases with host star metallicity \citep{2005ApJ...622.1102F, 2010PASP..122..905J}.

On the other hand, statistical analysis have revealed that $33-50$ \% of all stars host super-Earths within a 100 day period \citep{2013ApJ...766...81F}. This makes super-Earths the most abundant class of planets. The formation of these super-Earths, however, is still a mystery as not only their masses and orbital distances have to be matched by theoretical predictions, but also their period ratios. Recently, \citet{2017MNRAS.470.1750I} proposed that resonant chains formed by migrating super-Earths (also originating from beyond the snow line) have the potential to explain the observations of the period ratios of super Earth systems. For this, about $90$\% of the formed resonant chains have to become unstable and the resulting systems result in a very good match to the observations. On the other hand, whether the original planetary seeds form inside or outside the water ice line is still under debate.

In the core accretion model of planet formation, the planetary core of giant planets forms first. Only if the core is big enough it can start to accrete gas onto it to eventually form a giant planet \citep{1996Icar..124...62P}. Building planetary cores at large orbital distances with planetesimals alone can take longer than the typical lifetime of protoplanetary discs. Additionally, gravitational stirring of the planetesimals by a set of growing protoplanets makes the growth rates even lower \citep{2010AJ....139.1297L}. However, by taking the accretion of small mm-cm sized pebbles into account, the growth time-scale of planetary cores can be greatly reduced \citep{2010MNRAS.404..475J, 2010A&A...520A..43O, 2012A&A...544A..32L, 2012A&A...546A..18M}. These high growth rates, in turn, allow planets to grow locally before they start to migrate through the disc and eventually cross ice lines; this implies that their composition is based on the local composition of the accreted material.

The starting location of planetary seeds determines  the final orbital position and mass of growing planets where the pebble accretion scenario allows a local growth to a few Earth masses before the planets start to migrate towards the central star \citep{2015A&A...582A.112B, 2016A&A...590A.101B, BJ2017, 2017arXiv171010863N}. Therefore, constraining the starting position of planetary seeds that will either form giant planets or super-Earths will be of great help to constrain the pathways of forming planets and to refine planet formation theories.

One way to disentangle the formation location of planets is their chemical composition. The observations of atmospheres of hot Jupiters can reveal their chemical composition \citep{2017AJ....153...83B}, which can then be linked to their formation location \citep{2017MNRAS.469.4102M, 2017MNRAS.469.3994B}. However, these observations suffer from large uncertainties in the observational data and their interpretation. Additionally, observations can only probe the top layers of the planetary atmospheres, which might not necessarily be related to the bulk composition of the planet. On the theoretical side, there are many unknowns in the formation process of hot Jupiters regarding migration, core formation and gas accretion rates, making comparisons with observations very hard.

On the other hand, super-Earths and ice giants consist mostly of solids. A super-Earth forming in the hot inner parts of the protoplanetary disc will most likely contain very little water, while a super-Earth forming at $r>r_{\rm H_2O}$ will contain a significant fraction of water. The bulk density of super-Earths can be inferred if the super-Earth can be observed via transits and radial velocity at the same time. This reveals the radius and the mass of the planet, allowing the calculation of a mean density implying constraints on the planetary composition. However, these data have still large error bars \citep{2016AJ....152..160B} and the inferred composition is degenerate (e.g. \citealt{2010ApJ...712..974R,2017A&A...597A..37D}).

Another way to distinguish the formation location of giant planets is their imprint on the stellar abundances of their host star. This is very important in binary star systems, where both stars form from the same giant molecular cloud and thus presumably have the same chemical composition. The newly formed stars are surrounded by protoplanetary discs, which accrete onto them. During the formation of planets, the planets take out solid material from the protoplanetary disc, which is then not accreted onto the central star. If one star hosts a planet, while the other star does not, a chemical abundance difference between the stars in the binary system could be observed. These difference have already been observed and can be used to predict whether giant planets formed inside or outside the water ice line \citep{2014ApJ...790L..25T, 2015ApJ...808...13R, 2016ApJ...819...19T} even though the observations are not always showing clear abundance trends that allow predictions on the formation location of planets \citep{2014MNRAS.442L..51L, 2015A&A...582A..17S, 2016AJ....152..167T}.

\citet{2004A&A...420..683D, 2006A&A...454..581D} reported differences in [Fe/H] to be $\sim$0.03-0.1 dex for several pairs of binaries with stellar spectroscopy. Recent high-precision spectroscopic studies on planet hosting binaries have shown varied results. Abundance differences in binaries with trends in condensation temperature of elements ($T_{\rm cond}$) have been reported in several cases, e.g. XO-2 N+S \citep{2015ApJ...808...13R} and 16 Cygni A+B \citep{2014ApJ...790L..25T}. Abundance differences of refractories and volatiles in binaries with no significant $T_{\rm cond}$ trend have been reported by \citet{2016ApJ...819...19T, 2016AJ....152..167T} for WASP-94 and HD 133131A+B, respectively. \citet{2014ApJ...787...98M} reported abundance differences to be $\sim$-0.02 dex and -0.12 dex for C and O, respectively while $\sim$+0.05 dex for refractory elements in HD 20781/20782, although the uncertainties are large ($\sim$0.07 dex). \citet{2018ApJ...854..138O} reported an abundance difference to be $\sim$0.1-0.2 dex in co-moving solar-type binaries, although the difference is more likely due to the enrichment of planetary material. However, abundance differences do not always exist in binaries with giant planets. For example, \citet{2014MNRAS.442L..51L} and \citet{2015A&A...582A..17S} found no abundance difference in the HAT-P-1 system and in HD 80606/80607 within errors, respectively. The complexity of the observational results might be due to the different planets these systems host.

Compared to solar twins, the Sun's photosphere is depleted in refractory elements \citep{2009ApJ...704L..66M}. \citet{2010ApJ...724...92C} showed that adding 4 Earth masses of Earth-like and carbonaceous-chondrite-like material to the Sun's convection zone results in a composition in line with the mean value of solar twins. This implies that the observed solar composition could have arisen from a depletion of accreted material during the planet formation process. Similarly, \citet{2016MNRAS.456.2636L} showed that the Kepler-10 system is depleted in refractories compared to its twins, implying the possible imprinted chemical signatures related to terrestrial planet formation.

We can use a similar method to calculate the abundance difference on host stars caused by forming super-Earth and ice giants that take volatile and refractory material away from the protoplanetary disc which consequently can not be accreted any more by the central star. We assume in this work that the accretion process of the protoplanetary disc onto the star is the main driver of the disc's mass reduction in time. We propose here to use binary star systems as a method to distinguish observationally where super-Earth and ice giants, planets dominated by solids, formed in respect to ice lines of different molecular species. In particular to distinguish if super-Earths formed inside or outside the water ice line will help significantly to constrain planet formation theories.

However, planet formation in binary star systems is complicated compared to planet formation in single star systems, because the gravitational interactions of a close binary star can prevent the formation of planetesimals due to the generated large impact velocities between the planetesimals \citep{2013A&A...553A..71M}. Additionally, in close binaries (separation of a few 10 AU), stellar heating will inhibit dust coagulation, because the temperatures rise above the vaporization temperatures of many volatile materials \citep{2000ApJ...537L..65N}. Protoplanetary discs in close binary star systems can exchange material, influencing the star's composition, making close binaries not applicable for our model, which relies on independent evolution of the chemical composition of the stars after disc formation. Our model is therefore applicable to wide binary stars (a few 100 AU separation), where the protoplanetary discs do not exchange material and the heating of the neighbouring star is small enough that volatiles in the disc do not evaporate due to external heating processes and thus planet formation is indistinguishable from single stars \citep{2010ApJ...709L.114D}.

Our model is independent on the exact formation process of the super-Earth and ice giant, because the chemical composition of planetesimals and pebbles depends directly on their formation location. Planetesimals formed in the very inner disc will not contain water, while planetesimals formed at $r>r_{\rm H_2O}$ will contain water. The same applies for pebbles drifting through the disc: they will lose their water component when crossing the water ice line. Therefore, the resulting abundance difference in host stars is solely determined where the planetary building material comes from, but independent of the growth process of the planets (pebbles and/or planetesimals). However, we assume a fast local growth, so that the planet does not cross ice lines during its formation, but only starts migrating after solid accretion is complete, in agreement with the pebble accretion scenario \citep{2015A&A...582A.112B, 2016A&A...590A.101B, 2018A&A...609C...2B}. Our model will thus be able to allow observations to constrain the migration history of super-Earths and ice giant, independent of their exact formation processes (pebbles and/or planetesimals) and also independent of the exact disc structure, which influences the growth and migration of forming super-Earths \citep{2016A&A...590A.101B, 2017AJ....153..222B}.

Our work is structured as follows. In section~\ref{sec:methods} we describe our model to calculate the chemical composition of forming planets and how we calculate the predicted abundance difference of a star caused by the forming planet. In section~\ref{sec:results} we show the abundance differences in stars caused by planets formed at different ice lines and discuss the implications of different host star metallicities. Afterwards we discuss how to disentangle the formation location of planets from their imprint on the stellar abundances and show the effects that multiple planets and migrating planets have on our results in section~\ref{sec:location}. We discuss caveats of our model and implications of our results in section~\ref{sec:discuss} and we summarize in section~\ref{sec:summary}.

\section{Methods}
\label{sec:methods}

In order to calculate the abundance difference of stars with and without super-Earths or ice giants, we focus only on the solid material that is taken out of the protoplanetary disc from the forming planet. Therefore our model does not depend on the accretion process of the planet in itself, but only on the chemical model used. Even though super-Earths and also the ice giants in our own solar system have small atmospheres, we do not take their contribution to the stellar abundance difference into account. In section~\ref{sec:discuss} we also discuss how our model is influenced if the planets harbour small atmospheres.

Recent simulations have shown that the accretion of the protoplanetary disc is aided by disc winds launched at the surface of the disc \citep{2015ApJ...801...84G, 2016ApJ...818..152B, 2016ApJ...821...80B, 2016arXiv160900437S}. These winds carry away angular momentum from the disc, allowing the accretion of disc material onto the central star. However, the exact amount of material carried away from the disc and the location from where it leaves the disc is still under investigation.

In the late stages of disc evolution, photoevaporation can clear the disc in a very short time-scale and blow material away from the disc and the star (see \citealt{2013arXiv1311.1819A} for a review). Nevertheless, this process is very efficient only in the last few 100 kyr, when the formation of super Earths is presumably already finished. Additionally, this effect would probably carry away equal amounts of material from each disc in binary systems.

Nevertheless, in our work we assume that the protoplanetary disc in which planets form is completely accreted onto the central star. We first describe the chemical model used to calculate the solid ratios of the material transformed into planets and then how we calculate the stellar abundance difference.

The chemical composition of a star is determined by calculating abundances from spectral lines. The most common way to measure the abundance of an element X in the star is as a ratio to the hydrogen abundance H, normalized to the Sun, 
\begin{equation}
{\rm[X/H]} = \log_{10} \left( \frac{N_{\rm X}}{N_{\rm H}} \right)_{\rm star} - \log_{10} \left( \frac{N_{\rm X}}{N_{\rm H}} \right)_{\rm sun}
\label{eq:abundance}
\end{equation}
where $N_{\rm X}$ and $N_{\rm H}$ are the number of atoms per unit volume for element X and hydrogen, $N_{\rm X}/N_{\rm H}$ will be denoted as X/H. Note here that the measured stellar abundances reflect only the abundances in the outer stellar convective zone, which we assume to be $2.5$ \% ${\rm M}_\odot$ \citep{2010ApJ...724...92C}. However, \citet{2017A&A...599A..49K} showed that the HR diagrams of stellar clusters could be evidence that stellar evolution is not too far from classical models and that it might thus be unlikely that the star moves out of its pre-main sequence rapidly enough while the gas disc is still present as assumed in \citet{2010ApJ...724...92C}. This could imply that the effects of early planetesimal and planet formation might be overestimated, however, it does not affect abundance ratios between binary stars as discussed here. Nevertheless, our model takes different sizes of stellar convective zones into account.

\subsection{Chemical models}
\label{chemmodel}

At different temperatures in the disc, different molecular species condense out to solid form and can form pebbles. These pebbles can then form planetesimals, for example through the streaming instability (see \citealt{Johansen2014} for a review). The planetesimals and/or the pebbles can then be accreted onto forming planets. In table~\ref{tab:species} we show the condensation temperatures and elemental abundances used in \citet{2017MNRAS.469.4102M}, which are adapted from \citet{2011ApJ...743L..16O}. We assume constant condensation temperature $T_{\rm cond}$ as the dependence of $T_{\rm cond}$ on pressure is marginal \citep{2009A&A...507.1671M}. The mixing ratios (by number) of the different species as a function of the elemental number ratios is denoted X/H and corresponds to the abundance of element X compared to hydrogen for solar abundances, which we take from \citet{2009ARA&A..47..481A} and are given as follows: He/H = 0.085; C/H = $2.7\times 10^{-4}$; N/H = $7.1\times 10^{-5}$; O/H = $4.9\times 10^{-4}$; Mg/H = $4.0\times 10^{-5}$; Si/H = $3.2\times 10^{-5}$; S/H = $1.3\times 10^{-5}$; Fe/H = $3.2\times 10^{-5}$. Our results are therefore strictly speaking only valid for stars with solar composition, however, the compositions in the calculations can be changed to retrieve results for stars with different composition by changing the corresponding X/H value from the original solar value.

The temperatures in the protoplanetary disc are at maximum a few 100 K outside of 0.5 AU \citep{2015A&A...575A..28B}, where the inner disc heating is dominated by viscous heating and the outer disc by stellar irradiation \citep{2013A&A...549A.124B}. In binary stars, external heating from the binary companion can heat the disc, however, in the case of wide binaries (separation larger than a few 100 AU), this effect does not play a role. Therefore, hydrogen is not in atomic form as inside a star, but in molecular form H$_2$ in the disc. This means in protoplanetary discs the elements are normalised to H$_2$, while they are normalised to hydrogen atoms H in the stellar atmosphere. Additionally we also have to take the effects of helium into account, where [He/H] = 0.085, which changes the relative abundances in the following way
\begin{equation}
  [{\rm X}/{\rm H}_2] = 2 [{\rm X}/{\rm H}] f_{\rm H_2} \ ,
\end{equation}
where $f_{\rm H_2}$ corresponds to the transformation of the solar abundances of helium to H$_2$ in the disc. The value of $f_{\rm H_2}$ is calculated in the following way
\begin{equation}
 f_{\rm H_2} = \frac{1}{ 2 \times [{\rm He/H}] + [{\rm H/H}]} = 0.85 \ ,
\end{equation}
where $2 \times [{\rm He/H}]$ corresponds to $[{\rm He/H_2}]$ and $[\rm{H/H}]=1$ is the hydrogen volume mixing ratio relative to hydrogen.

If the temperature of the protoplanetary disc is less than $T_{\rm CO} = 20$ K, then all chemical species in our model will be in solid form, allowing an accretion of all chemical species either via pebbles or planetesimals onto the forming planet. For $T>T_{\rm CO}$, CO will only be available in gaseous form and thus planetesimals will have formed without a CO component and drifting pebbles will have lost their CO component, implying that a planet forming at $T>T_{\rm CO}$ will contain less carbon and oxygen compared to a planet forming at $T<T_{\rm CO}$. The rest of the disc is then accreted onto the star, so we can calculate the abundance difference in the atmospheres of their host stars and distinguish different formation locations.

For the chemical composition of the disc, we use a prescription based on theoretical calculations of chemistry in H$_2$ rich environments \citep{2009A&A...501..383W, 2011ApJ...743..191M}, named in the following as chemical model 1, and chemical compositions based on observations of ice and gas in protoplanetary discs \citep{2003ARA&A..41..241D, 2006A&A...453L..47P, 2011ApJ...743L..16O}, named chemical model 2. The main differences between these is how the carbon abundance is divided between the different molecules in the disc. The first chemical model does not contain any pure carbon grains at all, which means that planets forming at a temperature larger than 70 K will not contain any carbon.

\begin{table*}
\centerline{\begin{tabular}{c|c|c|c}
\hline
Species (Y) & $T_{\text{cond}}$ {[}K{]} & Chemical model 1 & Chemical model 2 \\ \hline \hline
CO & 20  & 0.45 $\times$ C/H (0.9 $\times$ C/H for $T < 70$ K) & 0.65 $\times$ C/H \\
CH$_4$ & 30 &  0.45 $\times$ C/H (0 for $T < 70$ K) & 0\\
CO$_2$ & 70 & 0.1 $\times$ C/H & 0.15 $\times$ C/H   \\
H$_2$O & 150 & O/H - (3 $\times$ Si/H + CO/H + 2 $\times$ CO$_2$/H) & Same as model 1\\
C$_{\rm grain}$ & 500 & 0 & 0.2 $\times$ C/H\\
MgSiO$_3$& 1500 & Si/H & Same as model 1 \\  \hline 
\end{tabular}}
\caption[Condensation temperatures]{Condensation temperatures and volume mixing ratios of the chemical species, adopted from \citet{2011ApJ...743L..16O}.}
\label{tab:species}
\end{table*}

We extend our chemical model 1 and 2 to include Fe$_2$O$_3$, Fe$_3$O$_4$, FeS, NH$_3$ and Mg$_2$SiO$_4$ as shown in table~\ref{tab:species2}. For convenience, only the extension of model 1 can be seen in table \ref{tab:species2}, because this extension is adoptable for both models. We assume that the iron which is not bound with sulphur in FeS, is in a 50:50-ratio between Fe$_2$O$_3$ and Fe$_3$O$_4$. We also divide the silicon in a 75:25-ratio between Mg$_2$SiO$_4$ and MgSiO$_3$ respectively, to accompany all magnesium atoms.

\begin{table*}
\centering
\begin{tabular}{c|c|c}
\hline
Species (Y) & $T_{\text{cond}}$ {[}K{]} & $v_{\text{Y}}$ \\ \hline \hline
CO & 20  & 0.45 $\times$ C/H (0.9 $\times$ C/H for $T < 70$ K) \\[5pt]
CH$_4$ & 30 & 0.45 $\times$ C/H (0 for $T < 70$ K)  \\[5pt]
CO$_2$ & 70 & 0.1 $\times$ C/H  \\[5pt]
NH$_3$ & 90* & N/H  \\[5pt]
H$_2$O & 150 & O/H - (3 $\times$ MgSiO$_3$/H + 4 $\times$ Mg$_2$SiO$_4$/H + CO/H \\
& & + 2 $\times$ CO$_2$/H + 3 $\times$ Fe$_2$O$_3$/H + 4 $\times$ Fe$_3$O$_4$/H) \\[5pt]
Fe$_3$O$_4$ & 371 & (1/6) $\times$ (Fe/H - S/H) \\[5pt]
C (carbon grains) & 500 & 0 \\[5pt]
FeS & 704 & S/H \\[5pt]
Mg$_2$SiO$_4$ & 1354 & 0.75 $\times$ Si/H  \\ [5pt]
Fe$_2$O$_3$ & 1357** & 0.25 $\times$ (Fe/H - S/H) \\ [5pt]
MgSiO$_3$ & 1500 & 0.25 $\times$ Si/H  \\  \hline
\end{tabular}
\caption[Condensation temperatures]{Condensation temperatures and volume mixing ratios of the chemical species. Condensation temperatures for added molecules from \citet{2003ApJ...591.1220L}. *Condensation temperature for NH$_3$ from \citet{2015A&A...574A.138T}.  **$T_{\text{cond}}$ for Fe$_2$O$_3$ the condensation temperature for pure iron is adopted \citep{2003ApJ...591.1220L}. Volume mixing ratios (i.e. by number) adopted for the species as a function of disc elemental abundances under two different prescriptions for condensate chemistry (see e.g. \citealt{2014ApJ...791L...9M}).}
\label{tab:species2}
\end{table*}

We note that volatile chemistry in protoplanetary discs can be extremely complex and depends on a number of parameters (e.g. \citet{2013ChRv..113.9016H}), especially when the abundance of complex molecules are of interest. Additionally, such discs can be significantly out of chemical equilibrium \citep{2011A&A...534A.132V}. Nevertheless, our model accounts for the prominent molecules dominating the O and C reservoirs (H$_2$O, CO$_2$, CO), while still rendering the problem tractable.

\subsection{Total amount of heavy elements in a planet}

The total amount of heavy elements found in a planet can be material directly bound in the planetary core during the core formation stage, but also material that is ablated and mixed into the primordial atmosphere of the formed planet \citep{2011MNRAS.416.1419H, 2016A&A...596A..90V, 2017ApJ...836..227L}. Here we assume that material in the primordial atmosphere can only be accreted when the planet accretes solids in general, like in the pebble accretion scenario, where pebbles can evaporate in the atmosphere of the planet, enhancing the heavy element content in the atmosphere \citep{2017arXiv170805392B}. To calculate how much mass of a molecular species Y (e.g. CO, CO$_2$, etc.) is found in a formed planet with total mass $M_{\rm P, tot}$, the ratios of the different molecular species have to be determined
\begin{equation}
 M_{\rm P, Y} = \frac{m_{\rm Y} v_{\rm Y}}{\sum (m_{\rm Y} v_{\rm Y})} M_{\rm P, tot} \ .
\end{equation}
We define $m_{\rm Y}$ as the molar mass for a given molecular species $Y$ (e.g. $m_{\rm CO} = 28$ g/mol), $v_{\rm Y}$ as the volume mixing ratio of specific molecules (see table~\ref{tab:species2}) and $\sum (m_{\rm Y} v_{\rm Y})$ is the summation over all molecular species that are available in solid form at the temperature of the disc where the planet forms (e.g. at $T=25$ K, all molecules except CO are in solid form and thus $\sum (m_{\rm Y} v_{\rm Y})$ does not contain any contribution of CO). Here the volume mixing ratios for the disc are used ([X/H$_2$]).

From the mass of each molecular species Y, we calculate the mass of each element X, which is needed to calculate the change of stellar abundances. For example, carbon is bound in CO, CO$_2$ and carbon grains C$_{\rm grain}$ for the second chemical model. By taking the ratios of the atomic masses of carbon (12 g/mol) in CO (28 g/mol), etc., we calculate the total mass of carbon in the planet $M_{\rm P,C}$ in the following way
\begin{equation}
 M_{\rm P, C} = \frac{12}{28} M_{\rm P, CO} + \frac{12}{44} M_{\rm P, CO_2} + M_{\rm P, C_{\rm grain}} \ .
\end{equation}

\subsection{Stellar abundances}
\label{subsec:diff}

The forming star is surrounded by a protoplanetary disc, which accretes onto the central star. If there is no other process that takes away solids from the protoplanetary disc, we assume that the solid grains accrete onto the star as well. Additionally, we can assume that volatiles that evaporate in the disc are accreted on the star as well with the bulk gas consisting of hydrogen and helium. This implies that the stellar abundance does not change when all the material from the disc is accreted onto the star. On the other hand, if a planet forms in the disc, it takes away solid material from the disc, which will lead to a change in the stellar abundance compared to the other binary star without planets. The change of the stellar abundance due to a planet taking material away from the disc should also depend on the mass of the original protoplanetary disc. Additionally it depends on the mass of the outer convective zone of the star, because this determines how much of the material of the accreted disc is mixed with stellar material that is then observable in the stellar atmosphere.

We calculate the change of the stellar abundance of a given chemical element X in the following way
\begin{eqnarray}
\label{eq:deltaXH}
 \Delta [{\rm X}/{\rm H}] &=& \log_{10} \Big( \frac{(M_{\rm CZ, X} - M_{\rm disc, X}) + (M_{\rm disc, X} - M_{\rm P, X})}{M_{\rm CZ, tot}} \nonumber \\ 
 &\times& \frac{\mu_{\rm Sun}}{m_{\rm X}} \frac{1}{[{\rm X}/{\rm H}]} \Big) \ .
\end{eqnarray}
Here, $M_{\rm CZ,X}$ denotes the mass of element X in the stellar convective zone, $M_{\rm D,X}$ denotes the mass of element X in the original protoplanetary disc, $M_{\rm P,X}$ is the mass of element X incorporated into the planet and $M_{\rm CZ, tot}$ represents the total mass of the stellar convective zone. The term $(M_{\rm CZ, X} - M_{\rm disc, X})$ denotes how much of element X of the convective zone is replaced by disc material and the term $(M_{\rm disc, X} - M_{\rm P, X})$ represents how much disc material of element X is taken away by the growing planet. This equation implies that the abundance difference $\Delta$[X/H] does not depend on the disc mass in itself, but only on the mass of the convective zone. Here, $M_{\rm disc, X}$ denotes the mass of element X inside the disc 
\begin{equation}
 M_{\rm disc, X} = \frac{ [{\rm X}/{\rm H_2}] m_{\rm X} M_{\rm disc, tot}}{\mu_{\rm disc}} \ .
\end{equation}
Here $M_{\rm CZ, tot}$ corresponds to the total mass of the convective zone, while $M_{\rm CZ, X}$ corresponds to the mass of element X inside the convective zone. We set for our standard calculations the mass of the convective zone to be $2.5$ \% ${\rm M}_\odot$ \citep{2010ApJ...724...92C}. The parameter $m_{\rm X}$ denotes the molecular weight of element X, and $\mu_{\rm disc} = 2.3$ and $\mu_{\rm Sun} = \mu_{\rm disc}  [{\rm X}/{\rm H}]/[{\rm X}/{\rm H_2}] = \mu_{\rm disc} \frac{1.17}{2} = 1.3455$ represent the mean molecular weight in the disc and in the star, respectively.  We note that $M_{\rm P, X}$ should never become larger than $M_{\rm disc, X}$, because then the planet would take more of the element X from the disc that is available. Additionally, eq.~\ref{eq:deltaXH} requires that the mass of the convective zone is at least as large as the mass of the disc. In case the disc mass is larger or equal to the mass of the convective zone, we can mathematically set $M_{\rm CZ} = M_{\rm disc}$, because the excess of mass of the disc would be hidden below the convective zone of the star, not influencing the abundances in the convective zone itself.

We note that the dust masses of protoplanetary discs can differ by more than an order of magnitude, even for discs around stars of the same mass and age \citep{2017AJ....153..240A}. Our assumption of a total disc mass of $2.5$ \% ${\rm M}_\odot$ corresponds to the higher end of disc masses in the survey by \citet{2017AJ....153..240A}, if a dust-to-gas ratio of 1\% is assumed. However, eq.~\ref{eq:deltaXH} shows that the mass of the disc has no influence on the relative change of abundance $\Delta$[X/H]. We discuss the implications of different disc masses and different sizes of the convective zone in section~\ref{subsec:convect}.

\section{Results}
\label{sec:results}

We present in this section the chemical compositions of planets that form in discs, where the temperature of the disc at the position where the planet forms is $200$, $100$, $50$ and $10$ K, respectively. This corresponds to planets that form at $r_{\rm P}<r_{\rm H_2O}$, $r_{\rm H_2O}<r_{\rm P}<r_{\rm CO_2}$, $r_{\rm CO_2}<r_{\rm P}<r_{\rm CO}$ and $r_{\rm P} > r_{\rm CO}$. We calculate the planetary composition for planets ranging from 1 ${\rm M}_{\rm E}$ to 20 ${\rm M}_{\rm E}$, as those are the typical mass ranges of super-Earths and ice giants. From the planetary composition, we calculate the change of stellar abundances (eq.~\ref{eq:deltaXH}), where we assume a solar metallicity and composition, unless stated otherwise.

In section~\ref{subsec:chem1} we present results in respect to the chemical model 1 and in section~\ref{subsec:chem2} with respect to chemical model 2. In section~\ref{subsec:metal} we discuss the influence of the stellar metallicity on our results. We discuss on how to distinguish the different planet formation locations in section~\ref{sec:location}. We use for the results presented here that $M_{\rm disc, tot} = M_{\rm C, tot} = $2.5$ \% {\rm M}_\odot$.

\subsection{Chemical model 1}
\label{subsec:chem1}

We discuss in this section the results of the chemical model 1, where no pure carbon grains are available, indicating that only for $T_{\rm disc} <$ 70 K carbon will appear in the planetary composition.

\subsubsection{Water ice line}

\begin{figure*}
 \centering
  \includegraphics[scale=0.55]{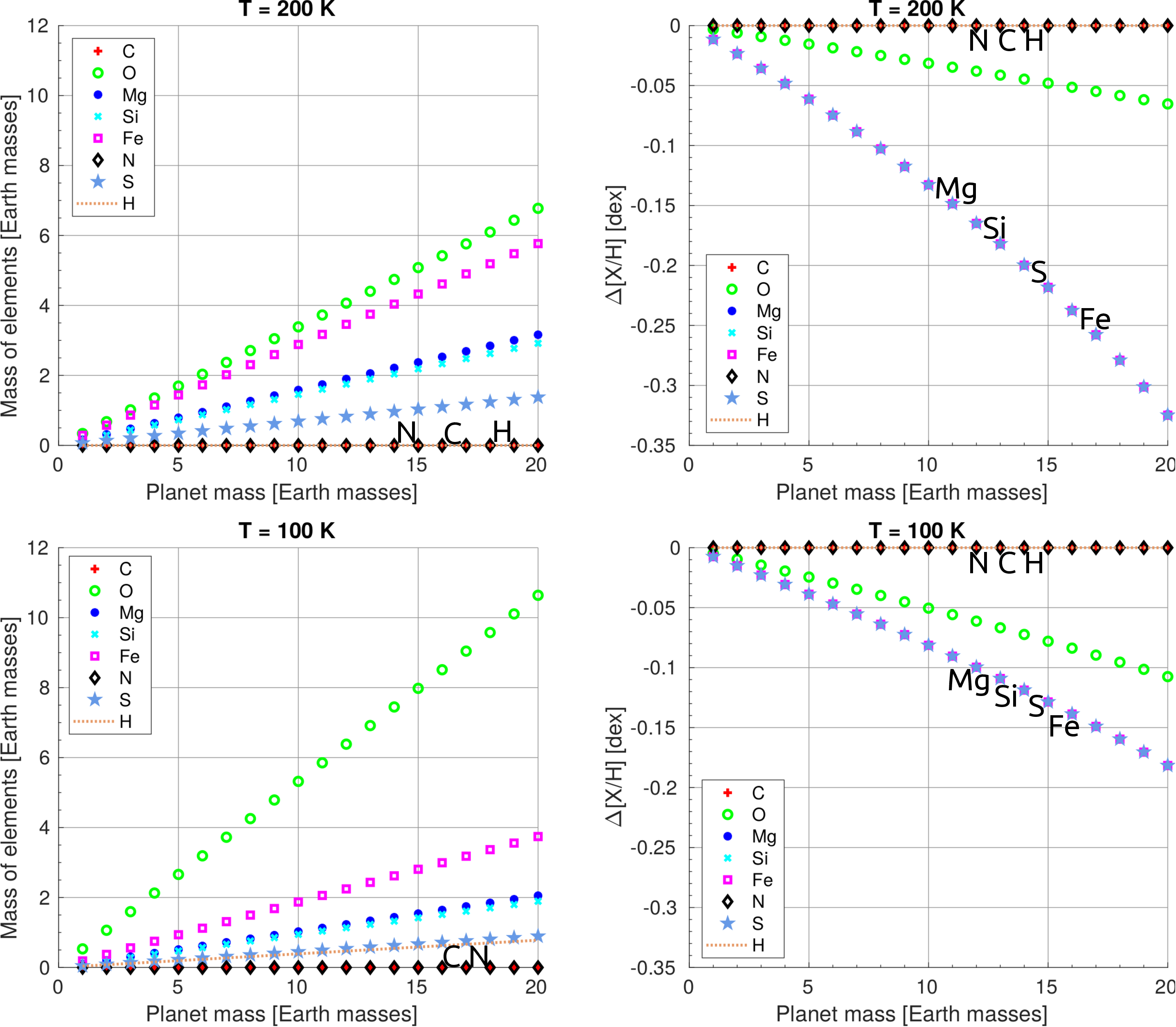} 
 \caption{Left: Elemental compositions of planets formed at $T_{\rm disc}=$ 200 K (top) and $T_{\rm disc}=$ 100 K (bottom) as function of planet mass. Right: Change of stellar abundances as function of planet mass for planets formed at $T_{\rm disc}=$ 200 K (top) and $T_{\rm disc}=$ 100 K (bottom). The lines for  iron, magnesium, silicon and sulphur all overlap for $\Delta$[X/H], because these elements are all accreted to a complete fraction into the planet and thus have the same relative abundance differences compared to a star without planets. We mark the chemical elements that show the same total mass in the planet and those that result in the same change of relative stellar abundance $\Delta$[X/H]. Here, chemical model 1 was used.
   \label{fig:H2Ochem1}
   }
\end{figure*}

In Fig.~\ref{fig:H2Ochem1} we show the planetary composition using the chemical species of table~\ref{tab:species2} for planets forming at $T_{\rm disc}=$ 200 K (top) and $T_{\rm disc}=$ 100 K (bottom), so for planets that form inside and outside of the water ice line, as function of the planetary mass. Additionally, we show the change of the stellar abundance as function of planetary mass.

On the left side in Fig.~\ref{fig:H2Ochem1} we show the planetary composition, which shows some difference between the two cases. The planet forming at 100 K shows a much larger oxygen abundance, because of the accretion of water ice. In turn, for a planet of the same mass this implies that all the other elements incorporated into the planet are reduced compared to the planet forming at 200 K. This is caused by the accreted water that makes up a significant fraction of the overall mass, which then results in a larger oxygen abundance for the planet forming at 100 K.

The right side of Fig.~\ref{fig:H2Ochem1} shows the change of the stellar abundances due to the material accreted onto the planet, which is then missing in the accreted protoplanetary disc, where the remainders are accreted onto the central star. Elements that are not in solid form at either 100 K or 200 K (e.g. all carbon bearing species) do not show a change in stellar abundance. However, most chemical species follow the same trend as a function of planet mass, except oxygen. The reason for that is that all iron, magnesium, silicon and sulphur bearing species are completely in solid form at T$\leq$ 200 K  and are thus accreted with a complete fraction into the planet in contrast to oxygen (e.g. the oxygen in CO is still in gaseous form and thus not accreted). The change of the stellar abundance of a specific element depends on how much of it is incorporated into the planet. If two or more elements are incorporated to a complete fraction into the planet (all elements with $T_{\rm c} >$ 200 K), the change of the stellar abundance is the same for all these elements, because the same relative fraction of these elements is taken away from the disc by the accreting planet.

Oxygen, on the other hand, is not accreted in a complete fraction in both cases, because some oxygen is bound in CO$_2$ and CO, which is not accreted in solid form. For the planet formed at 100 K, the oxygen abundance of the planet increases and thus also the change of stellar abundance for oxygen increases. On the other hand the change of stellar abundance reduces for all other elements compared to the planet formed at 200 K, because the planet contains less of these materials and thus takes away less material form the disc.

This is caused by the fact that at 100 K, also water ice is available, which is incorporated into the planet. This water ice then takes a significant fraction of the planetary mass, which results in a smaller relative abundance of all the other elements in the planet. As a consequence $\Delta$[X/H] for elements that are to 100\% in solid form at T$\leq$ 200 K is smaller for planets forming at T=100 K compared to planets forming at 200 K. The replacement of material that is to 100\% in solid form at T$\leq$ 200 K continues as the formation temperature of the planet decreases and more volatile species (H$_2$O, CO$_2$, CO, NH$_3$) become available at colder temperatures, reducing $\Delta$[X/H] for Si, Mg, S and Fe for even colder formation temperatures. This holds for all temperatures and both chemical models.

\subsubsection{CO ice line}

\begin{figure*}
 \centering
 \includegraphics[scale=0.55]{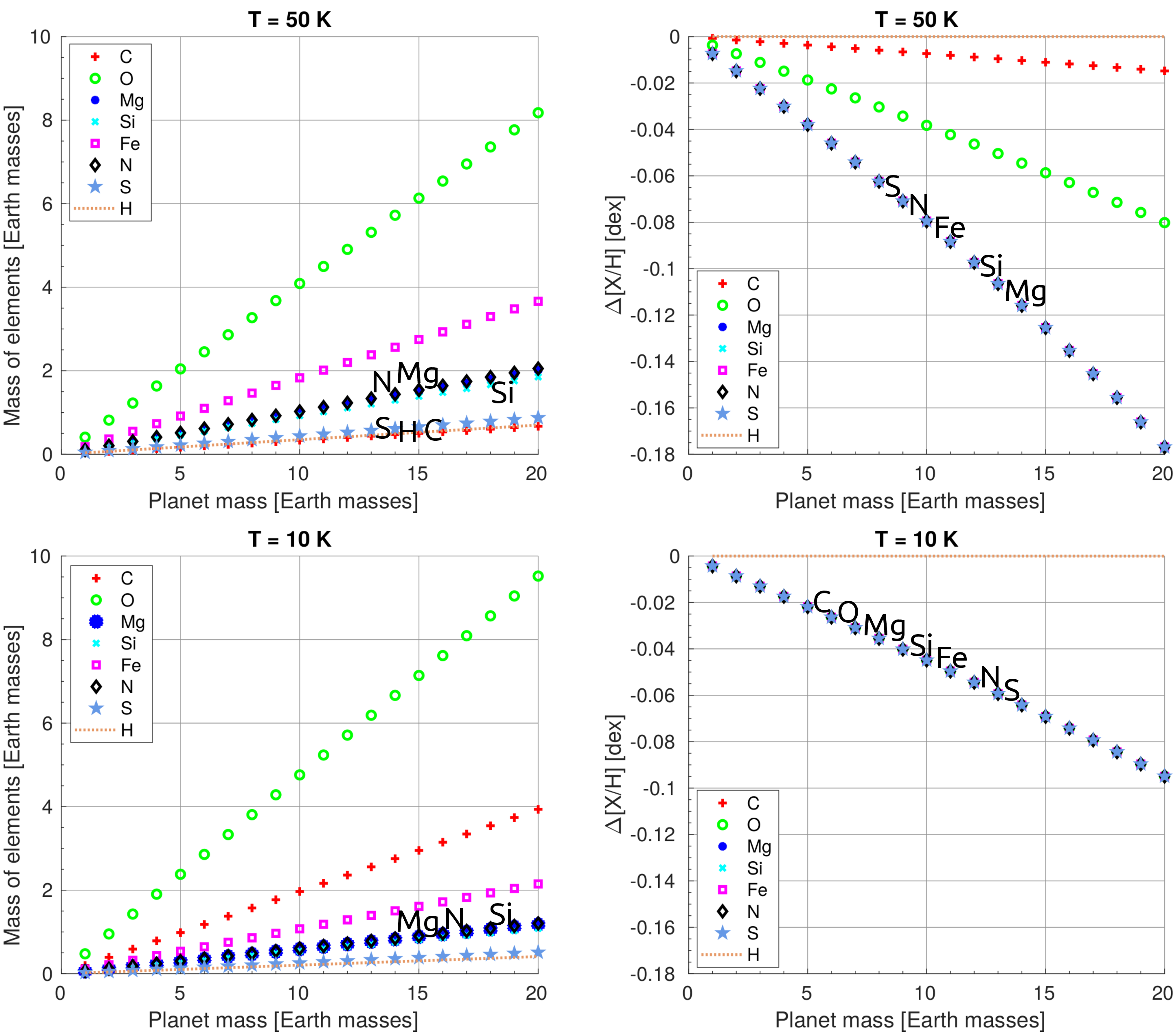}
 \caption{Left: Elemental compositions of planets formed at $T_{\rm disc}=$ 50 K (top) and $T_{\rm disc}=$ 10 K (bottom) as function of planet mass. Right: Change of stellar abundances as function of planet mass for planets formed at $T_{\rm disc}=$ 50 K (top) and $T_{\rm disc}=$ 10 K (bottom). The lines for  iron, magnesium, silicon, nitrogen and sulphur all overlap for $\Delta$[X/H], because these elements are all accreted completely into the planet and thus have the same relative abundance differences compared to a star without planets. In the bottom plot, all lines overlap, because all elements are in solid form and thus accreted to a complete fraction into the planet resulting in the same $\Delta$[X/H] value. Here, chemical model 1 was used.
   \label{fig:COchem1}
   }
\end{figure*}

In Fig.~\ref{fig:COchem1} we show the chemical composition of planets formed at $T_{\rm disc}=$ 50 K (top) and $T_{\rm disc}=$ 10 K (bottom) and the change of stellar abundance for the same planets. At these temperatures, nitrogen becomes available in solid form for the first time.

The planet formed at a disc temperature of $T_{\rm disc}=$ 50 K shows less oxygen abundance compared to the planet formed at $T_{\rm disc}=$ 100 K, even though CO$_2$ is now available in solid form, which should increase the oxygen abundance. However, in chemical model 1, the abundance of CO changes, when the temperature drops below 70 K. For these cold temperatures carbon is more efficient in forming CO than CH$_3$, so that a lot of oxygen gets locked in CO, which is not available in solid form at 50 K. However, as more oxygen is locked in CO, less water ice is available for the forming planet. This reduces the amount of oxygen in the planet compared to a planet forming at 100 K. Additionally, the planet now contains nitrogen and carbon. Now all chemical elements show a signature in the stellar abundance trend, where carbon and oxygen do not show a trend as strong as all the other elements, because not all carbon and oxygen bearing species (like CO) are in solid form.

At 10 K, all chemical molecules in our model are in solid form. This implies that the planet now contains a significant fraction of carbon in contrast to higher formation temperatures. Additionally, this implies that the change of the stellar abundances is the same for all chemical elements, because they are all accreted to a complete fraction. Note here that even though the planetary composition includes hydrogen, it can not be observed in the change of stellar abundances.

\subsubsection{Condensation temperature}

\begin{figure}
 \centering
 \includegraphics[scale=0.45]{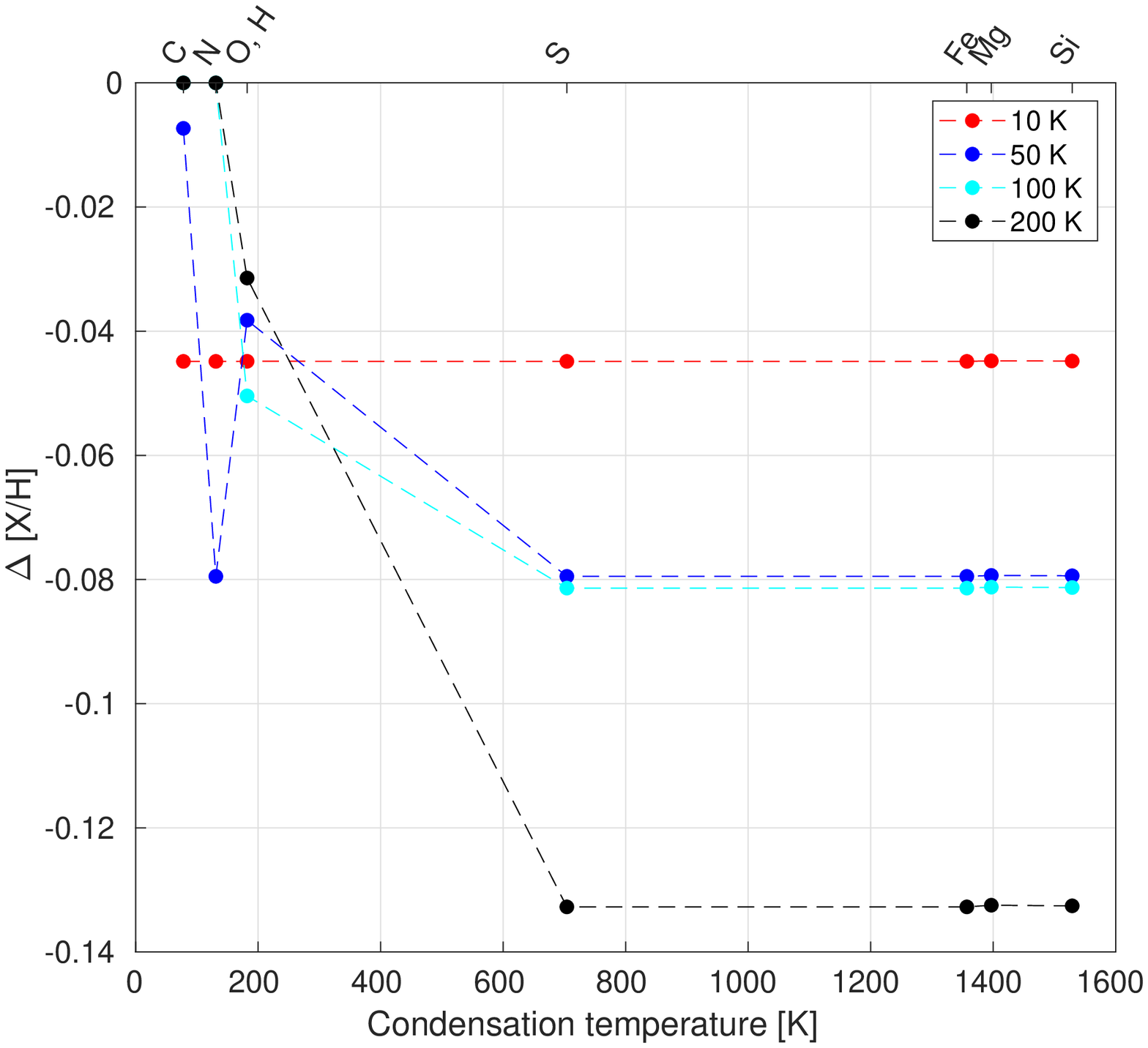} 
 \caption{Change of stellar abundance of all chemical elements X as a function of condensation temperature, where the condensation temperature for each element X is listed in the box in the top right corner, for 10 ${\rm M}_{\rm E}$ planets using chemical model 1. The different colours correspond to the different temperatures at which the planet forms, resulting in different overall compositions of the planet and thus different abundance trends. The lines are to guide the eye to identify which points belong together. 
   \label{fig:Condchem1}
   }
\end{figure}

In order to compare more easily the changes of the stellar abundances for planets formed at different temperatures in the disc, we show in Fig.~\ref{fig:Condchem1} the change of stellar abundances of elements X as a function of their condensation temperature $T_{\rm cond}$ for 10 ${\rm M}_{\rm E}$ planets. As mentioned before, the elements sulphur, magnesium, silicon and iron show the same abundance differences for each planet, because those elements are incorporated to a complete fraction into the planet at all disc temperatures. This results in the same change of abundance difference $\Delta$[X/H] for all these elements for a given formation temperature of the planet.

Fig.~\ref{fig:Condchem1} also shows the change of the oxygen abundance difference for planets formed at different formation temperatures. The change in the oxygen abundance is largest for the planet formed at 100 K, because of the water ice that is incorporated into the planet. For even colder temperature, the oxygen abundance decreases, because of the chemical model (as explained above).

For all planets, except for the one forming in the very cold parts of the disc ($T =$ 10 K), a clear trend between refractories (Si, Mg, Fe) and volatiles (O, C, N) is visible. Refractories are depleted in the star's atmosphere much more compared to volatiles. This trend is very interesting, because \citet{2009ApJ...704L..66M} found that the Sun is depleted in refractories relative to volatiles when compared to the majority of solar twins without known planets. They interpreted this trend as a sign of terrestrial planet formation, in agreement with our model. In a similar way, \citet{2016MNRAS.456.2636L} studied the Kepler-10 system and compared the relative abundance difference to its stellar twins and found a depletion in refractories in Kepler-10 of the order of $\sim$0.02 dex. These abundance differences are smaller than shown in Fig.~\ref{fig:Condchem1} even though the two planets in the system are in total $\sim$ 20${\rm M}_{\rm E}$. The abundance difference between our model and the observations could be caused by the larger stellar convective zone in Kepler-10 (see section~\ref{subsec:convect} for a discussion about the influence of the convective zone). However, this trend vanishes for ice giants in our model, due to the large fraction of volatiles incorporated into their cores, which agrees with the finding by \citet{2009ApJ...704L..66M} that such a trend is more likely the chemical signatures imprinted by terrestrial planet formation rather than ice giant formation. Here, both, theory and observations, predict that the $T_{\rm cond}$-trend is very likely due to terrestrial planet formation rather than giant planet formation, which would also contain a lot of volatiles in their atmospheres due to gas accretion \citep{2017MNRAS.469.4102M, 2017MNRAS.469.3994B}.

\subsection{Chemical model 2}
\label{subsec:chem2}

We discuss here the results of our simulations with chemical model 2, which features pure carbon grains in the disc in contrast to chemical model 1.

\subsubsection{Water ice line}

\begin{figure*}
 \centering
 \includegraphics[scale=0.55]{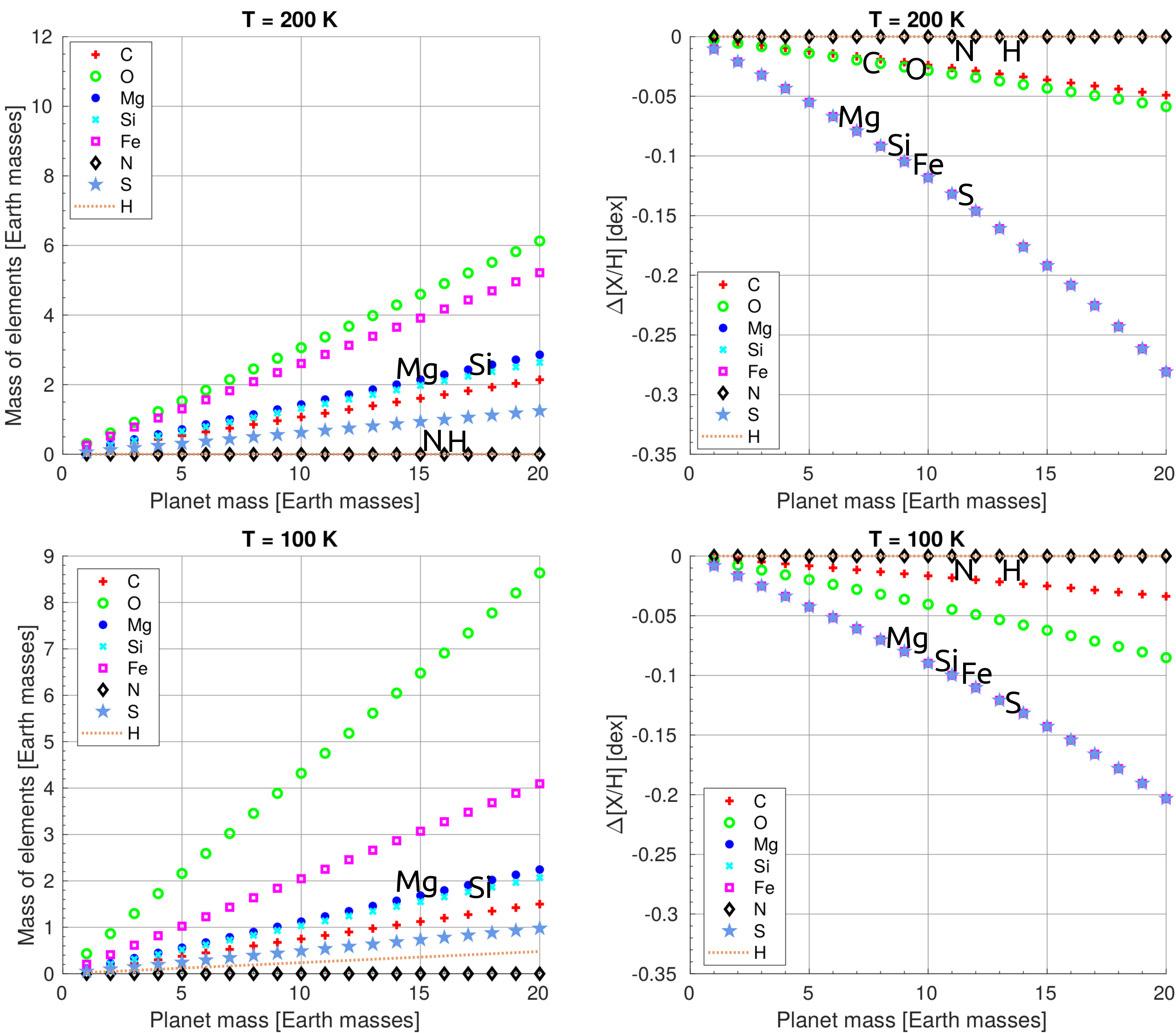} 
 \caption{Left: Elemental compositions of planets formed at $T_{\rm disc}=$ 200 K (top) and $T_{\rm disc}=$ 100 K (bottom) as function of planet mass. Right: Change of stellar abundances as function of planet mass for planets formed at $T_{\rm disc}=$ 200 K (top) and $T_{\rm disc}=$ 100 K (bottom). The lines for  iron, magnesium, silicon and sulphur all overlap for $\Delta$[X/H], because these elements are all accreted to a complete fraction into the planet and thus have the same relative abundance differences compared to a star without planets. Here, chemical model 2 was used.
   \label{fig:H2Ochem2}
   }
\end{figure*}

In Fig.~\ref{fig:H2Ochem2} we show the chemical composition of planets formed at $T_{\rm disc}=$ 200 K (top) and $T_{\rm disc}=$ 100 K (bottom) as a function of their mass. Additionally, the change of stellar abundance is shown. The main difference with respect to chemical model 1 is the inclusion of carbon grains in the calculations. This means that only nitrogen and hydrogen is not present in the planetary composition at $T_{\rm disc}=$ 200 K.

Similar to chemical model 1, the mass of oxygen of the planet increases for $T_{\rm disc}=$ 100 K due to the inclusion of water ice grains in the planet. However, this increase is not as pronounced as in chemical model 1, because a large fraction of oxygen is stored in CO, which is in gaseous form at those temperatures and can thus not be accreted. The change of stellar abundance due to the material taken out by the planet is shown on the right hand side in Fig.~\ref{fig:H2Ochem2}. The trends are similar as for chemical model 1.

\subsubsection{CO ice line}

\begin{figure*}
 \centering
 \includegraphics[scale=0.55]{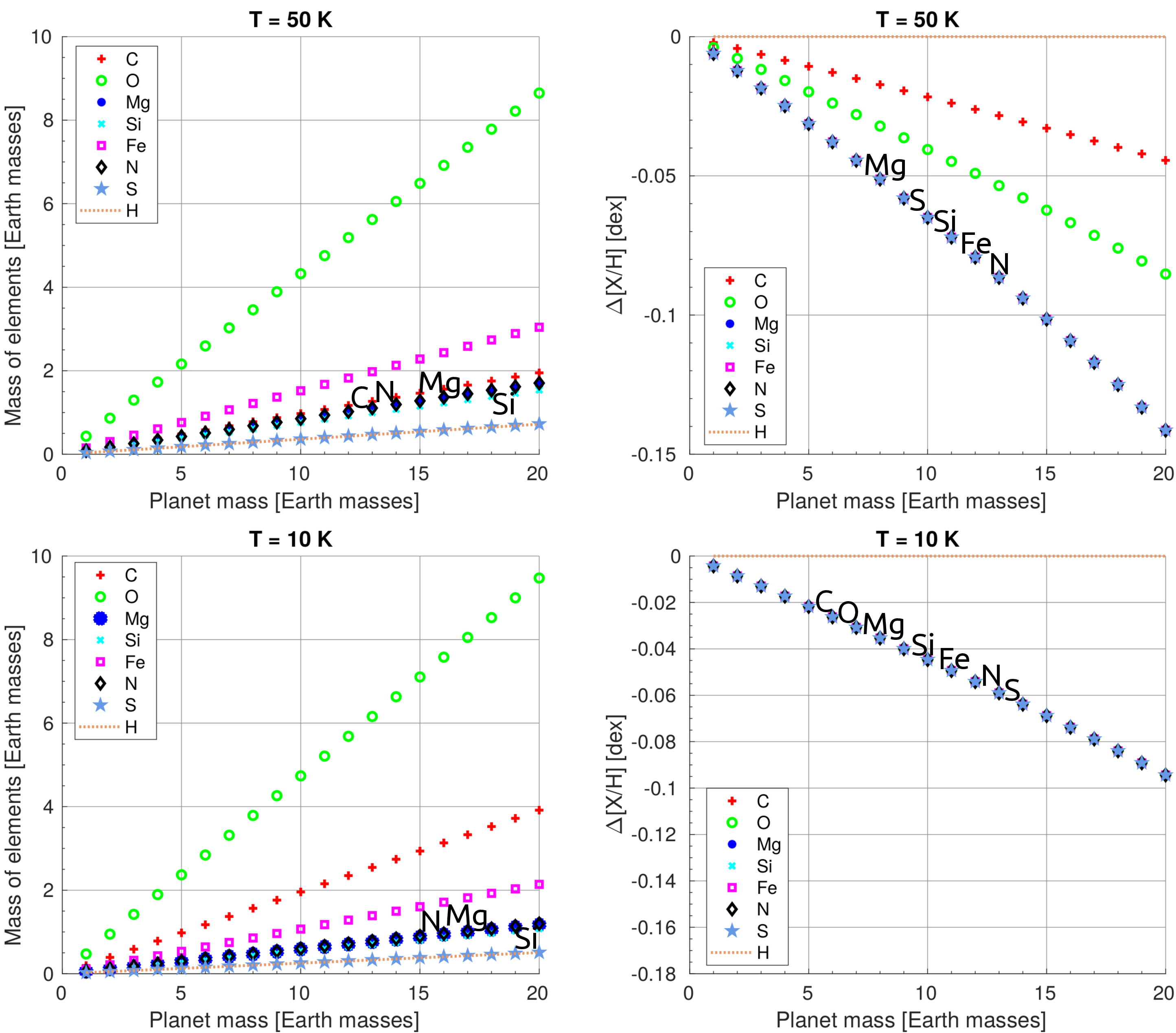} 
 \caption{Left: Elemental compositions of planets formed at $T_{\rm disc}=$ 50 K (top) and $T_{\rm disc}=$ 10 K (bottom) as function of planet mass. Right: Change of stellar abundances as function of planet mass for planets formed at $T_{\rm disc}=$ 50 K (top) and $T_{\rm disc}=$ 10 K (bottom).  The lines for  iron, magnesium, silicon, nitrogen and sulphur all overlap for $\Delta$[X/H], because these elements are all accreted to a complete fraction into the planet and thus have the same relative abundance differences compared to a star without planets. In the bottom plot, all lines overlap, as all elements are in solid form and thus accreted to a complete fraction into the planet. Here, chemical model 2 was used.
   \label{fig:COchem2}
   }
\end{figure*}

In Fig.~\ref{fig:COchem2} we show the chemical composition for planets formed in regions of the disc where $T_{\rm disc}=$ 50 K (top) and $T_{\rm disc}=$ 10 K during planet formation. Compared to $T_{\rm disc}=$ 100 K, the oxygen abundance increases slightly for $T_{\rm disc}=$ 50 K, while the oxygen abundance was reduced in this case for chemical model 1. Additionally, nitrogen appears in the chemical composition, because the sublimation temperature of NH$_3$ is crossed as well. The chemical compositions of the planet formed at $T_{\rm disc}=$ 10 K is exactly the same for both chemical models. This is caused by the fact that all volatile species included in both our chemical models are in solid form at $T_{\rm disc}=$ 10 K and thus the planets have the same composition.

In the right hand side of Fig.~\ref{fig:COchem2} we show the change of stellar abundance as a function of planetary mass for the different formation temperatures of the planets. Here now also nitrogen appears. As before, the change of stellar abundances is the same for all chemical elements X in the case of $T_{\rm disc}=$ 10 K.

\subsubsection{Condensation temperature}

\begin{figure}
 \centering
 \includegraphics[scale=0.45]{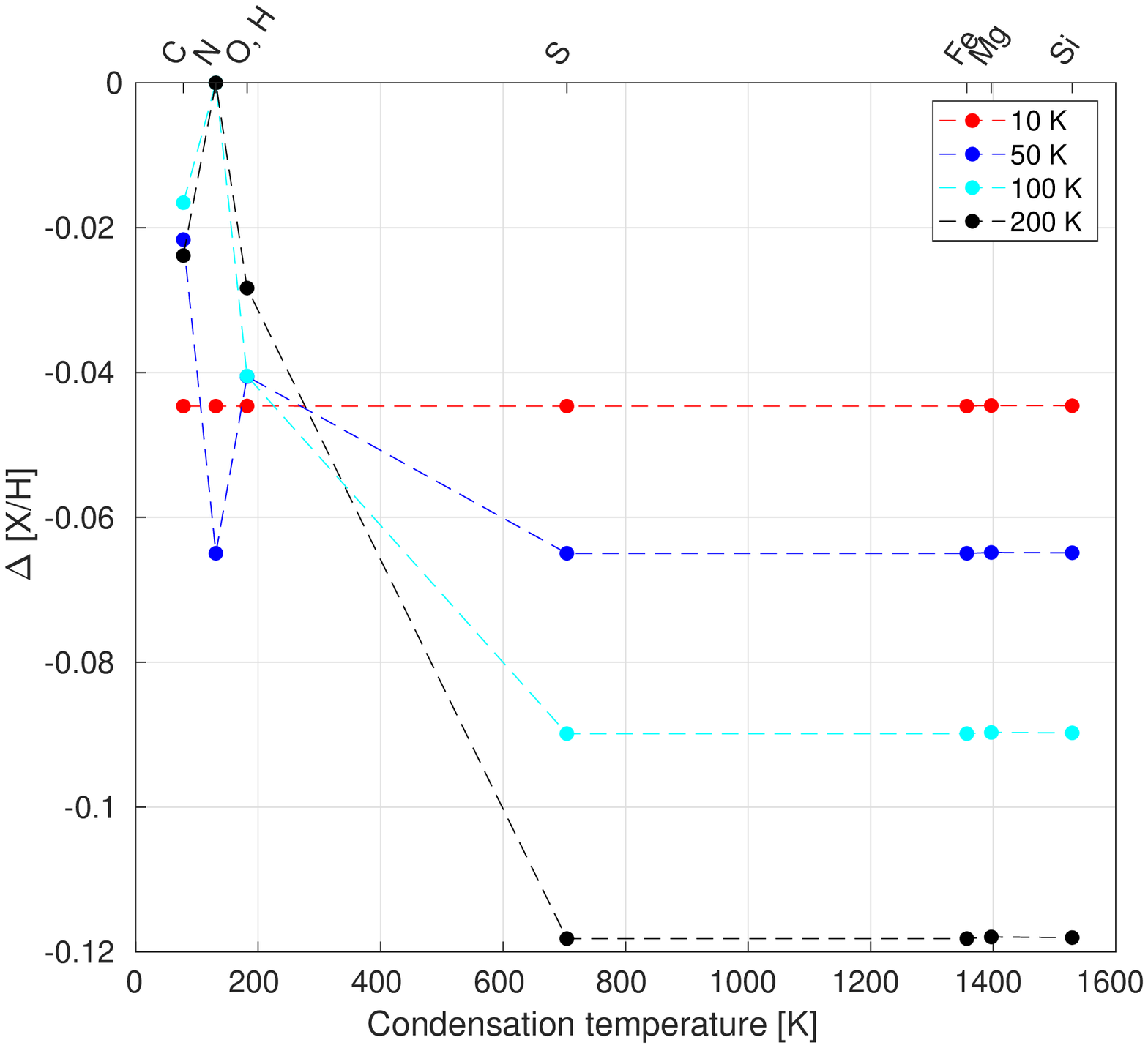} 
 \caption{Same as Fig.~\ref{fig:Condchem1}, but for chemical model 2.
   \label{fig:Condchem2}
   }
\end{figure}

In Fig.~\ref{fig:Condchem2} we show the stellar abundance difference for 10 ${\rm M}_{\rm E}$ planets formed at different disc temperatures as a function of the condensation temperatures of chemical elements X. As in chemical model 1, the elements sulphur, iron, magnesium and silicon have the same abundance differences for each planet, because all species are completely incorporated into planets independent of the formation temperature. However, in contrast to chemical model 1, there is a difference of about 0.03 dex for those elements between the planet forming at $T_{\rm disc}=$ 100 K and $T_{\rm disc}=$ 50 K.

\subsection{Change of stellar metallicity}
\label{subsec:metal}

The above calculations have been considering solar metallicity of the original parent molecular cloud that formed the binary star system. However, stars span a wide range of metallicities and also giant planet formation scales with host star metallicity \citep{2005ApJ...622.1102F}. If a 10 Earth mass planet forms around a star, the fraction of heavy elements it takes out from the protoplanetary disc compared to the amount of heavy elements of the disc itself is larger (lower) if the protoplanetary disc has lower (larger) metallicity. This changes the relative abundances that can be observed in host stars. For our calculations we therefore scale the mixing ratios (by number) of the different species as a function of the elemental number ratios of the solar composition X/H used before by the change in metallicity we want to study. This means that for [Fe/H]=0.2 systems, we increase the amount of all elements X by a factor of $10^{0.2}$ and recalculate the analysis described above for planets formed at different temperatures in the disc.

\begin{figure}
 \centering
 \includegraphics[scale=0.45]{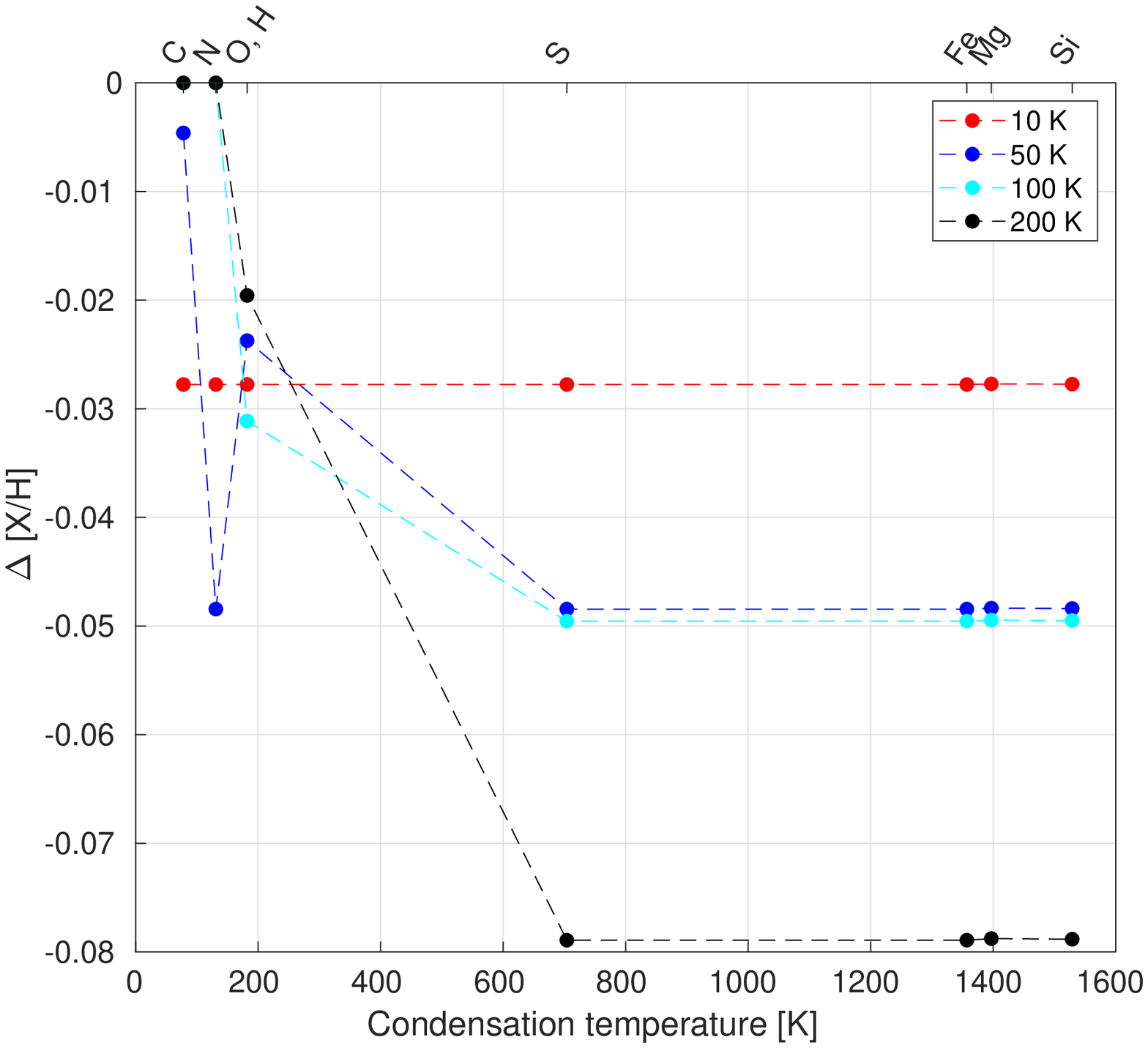} 
 \includegraphics[scale=0.45]{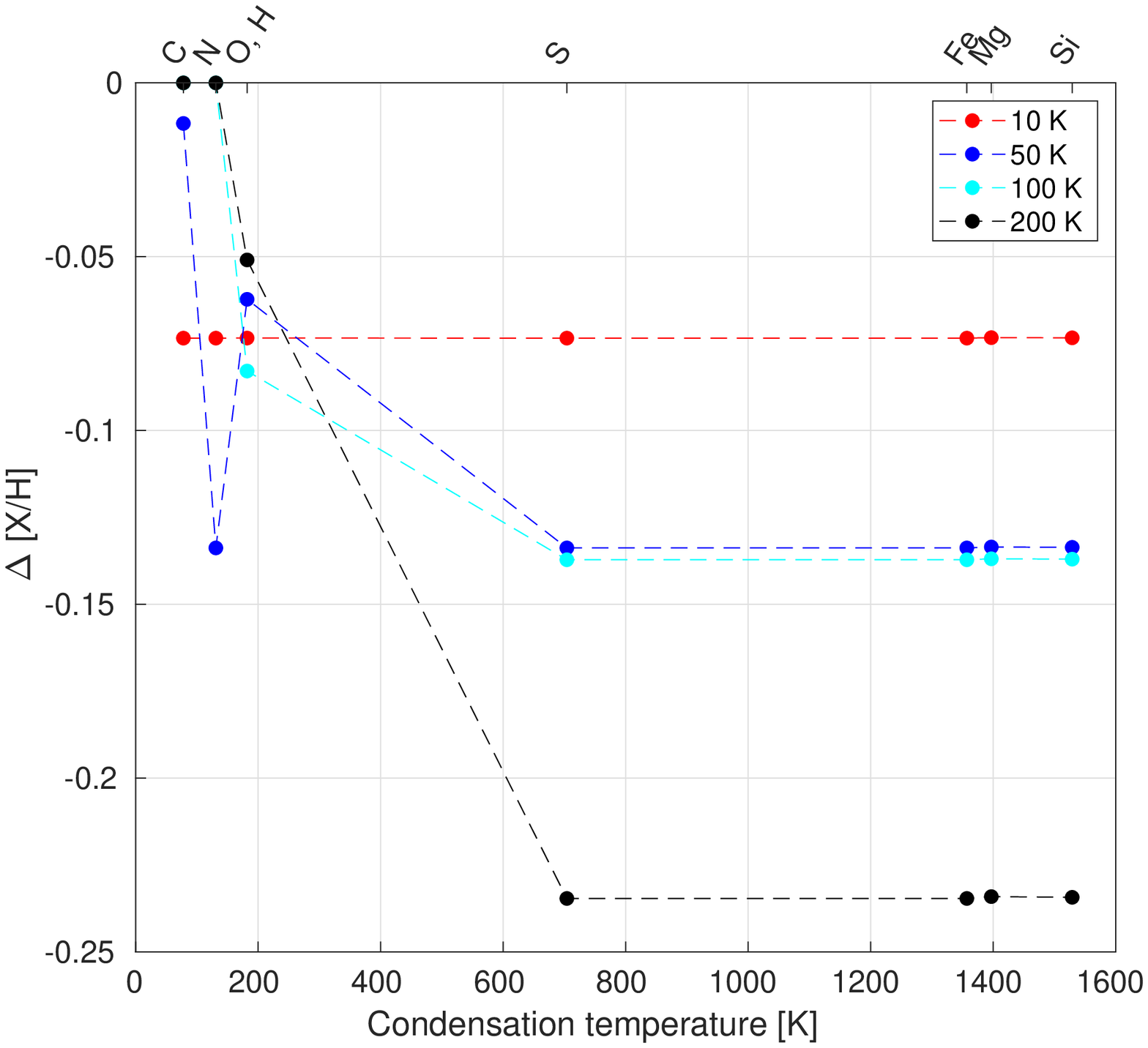}  
 \caption{Same as Fig.~\ref{fig:Condchem1}, but for [Fe/H]=0.2 (top) and [Fe/H]=-0.2 (bottom). Metal rich discs result in a smaller $\Delta$[X/H] compared to metal poor discs for the same planet mass. This makes looking for observable differences easier in metal poor systems.
   \label{fig:CondchemZ}
   }
\end{figure}

In Fig.~\ref{fig:CondchemZ} we show the change of the host star metallicity of element X as a function of the elemental condensation temperature for 10 ${\rm M}_{\rm E}$ planets using chemical model 1, where the top plot shows [Fe/H]=0.2 and the bottom plot [Fe/H]=-0.2. The changes of abundances follow the trends described above, where the sulphur, magnesium, silicon and iron abundance trends show all the same value for the different planet formation temperatures, because they are all accreted completely even at $T_{\rm disc}=200$ K.

The relative changes of the abundances of all elements in the host stars atmosphere are less pronounced in case the host star is metal rich ([Fe/H]=0.2) compared to a metal poor host star ([Fe/H]=-0.2). This allows observationally an easier distinction where a planet formed relative to an ice line for low metallicity stars. Additionally, a lower host star metallicity also allows to make distinctions for lower mass planets compared to a high metallicity host star, assuming the observational error is $\approx$0.01 dex. Additionally, metal poor stars are known to host less giant planets \citep{2005ApJ...622.1102F}, which reduces the probability to have giant planets in the system that can influence our predictions (see section~\ref{subsec:multi}), but at the same time the frequency of super-Earth planets remains unchanged, making low metallicity binaries ideal targets for future observations.

\section{Distinguishing the formation location of planets}
\label{sec:location}

In order to distinguish the formation location of planets, we compare the change in stellar abundances for different elements. In previous observations of binary star systems hosting planets, accuracies of [Fe/H]$\approx 0.01$ dex for the measurement of the Fe abundance have been obtained \citep{2014MNRAS.442L..51L, 2014ApJ...790L..25T, 2016ApJ...819...19T, 2016AJ....152..167T}. By looking at Fig.~\ref{fig:Condchem1} and Fig.~\ref{fig:Condchem2} we can already identify that just measuring the abundance differences of sulphur, silicon, magnesium and iron could help to distinguish between the formation location, because the differences in the stellar composition are larger than the error-bar of observations for the planets formed at different temperatures. However, here the convective zone of the central star might dilute predictions from just looking at single elements. We discuss the influence of the disc mass and of the convective stellar zone in section~\ref{subsec:convect}.

However, the difference depends on the chemical model used, where chemical model 1 allows to distinguish easier for planets formed inside or outside the water ice line. In the following we take a closer look on how to distinguish the formation location at different ice lines.

\subsection{Water ice line}

\begin{figure*}
 \centering
 \includegraphics[scale=0.7]{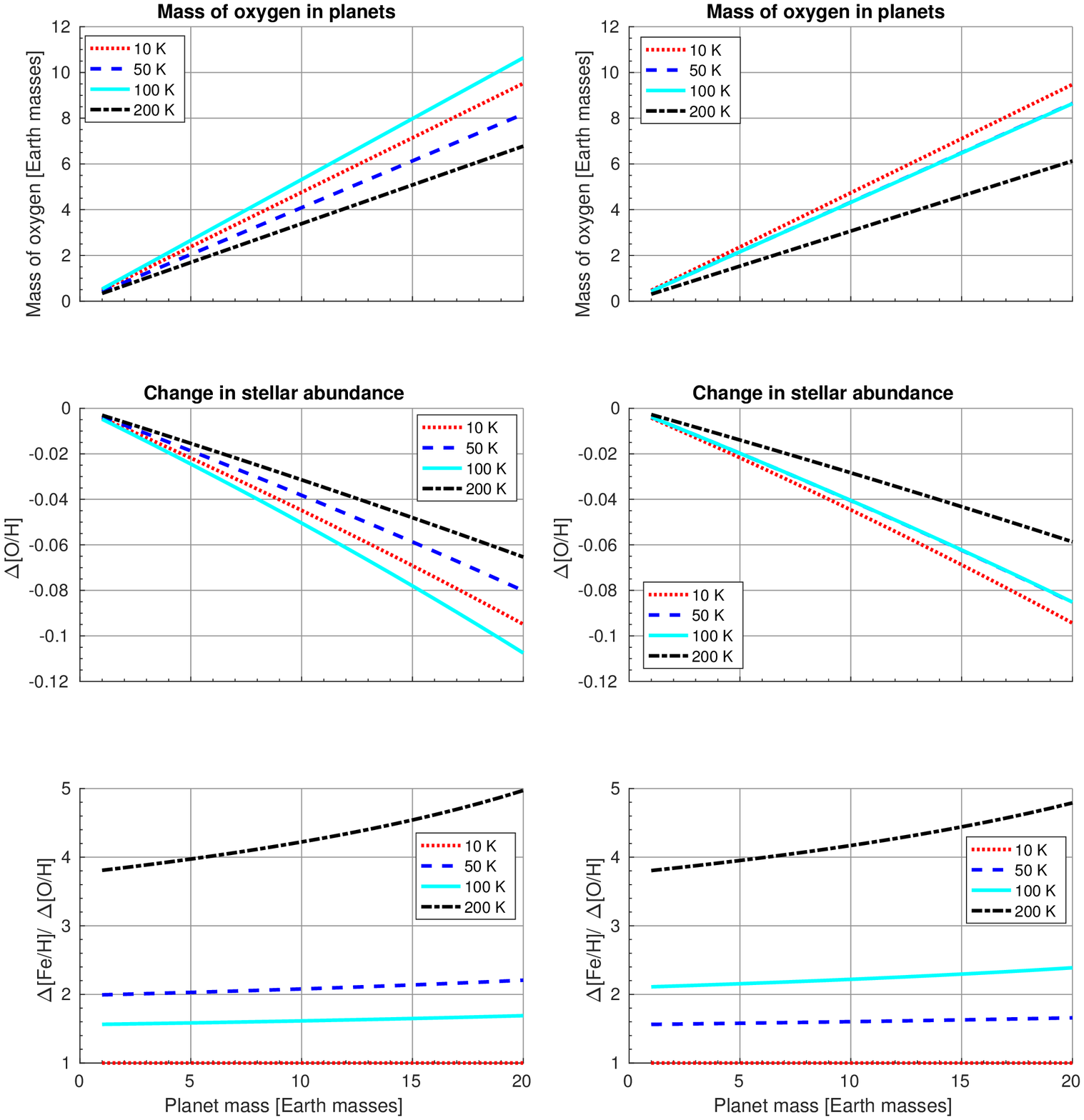} 
 \caption{Mass of oxygen in the planets as function of planetary mass (top). Oxygen abundance difference in the host star $\Delta$[O/H] as a function of planetary mass (middle). Ratio of the iron abundance difference $\Delta$[Fe/H] to the oxygen abundance difference $\Delta$[O/H] as function of planetary mass (bottom). All plots are for planets forming at different disc temperatures and for both chemical models, where chemical model 1 is on the left and chemical model 2 on the right.
   \label{fig:Ochem}
   }
\end{figure*}

Planets forming at $r>r_{\rm H_2O}$ contain a significant amount of water ice and thus oxygen. The increase of oxygen is slightly larger in chemical model 1 compared to chemical model 2, but the differences for the planets formed at different $T_{\rm disc}$ is still significant. In Fig.~\ref{fig:Ochem} we show the mass of oxygen in the planets (top), the stellar abundance difference of oxygen (middle) and the ratio of the iron abundance difference $\Delta$[Fe/H] to the oxygen abundance difference $\Delta$[O/H] (bottom) for both chemical models.

The change of the oxygen abundance compared to the different formation temperatures of the planets seems to allow to distinguish their formation location by the oxygen abundance only if the planets are more massive than 10 Earth masses, if one assumes observational errors of [Fe/H]$\approx$0.01 dex. However, $\Delta$[O/H] depends on the convective zone and the disc mass, which are essentially unknown and change the $\Delta$[O/H] value, making predictions if a planet formed inside or outside the water ice line not reliable by just looking at $\Delta$[O/H] alone.

However, combining the measurements of $\Delta$[O/H] with the measurements of $\Delta$[Fe/H] can reveal additional information. The ratio of the  iron abundance difference $\Delta$[Fe/H] to the oxygen abundance difference $\Delta$[O/H] is about a factor of 2 larger for planets forming at $r<r_{\rm H_2O}$ ($T_{\rm disc} = 200$ K model) compared to planets forming at $r>r_{\rm H_2O}$ ($T_{\rm disc} = 100$ K model), independent of the chemical model. By relating $\Delta$[Fe/H]/$\Delta$[O/H] to the formation location of planets, we actually eliminate the effects of the convective zone (see section~\ref{sec:discuss}).

We note that $\Delta$[Fe/H] could also be replaced by $\Delta$[S/H], $\Delta$[Si/H] or $\Delta$[Mg/H], because all these elements are accreted to a complete fraction at all formation temperatures in our model. The best approach for observations would thus be to measure the abundances of all those elements in the host star to minimize the error.

In chemical model 1, the planets have a larger oxygen content at 100 K compared to chemical model 2, while at 200 K, the oxygen content is roughly equal. Additionally, for both chemical models, the Fe content is roughly equal for 200 K and 100 K. This then leads to a reduced $\Delta$[Fe/H]/$\Delta$[O/H] value for chemical model 1 compared to chemical model 2 for the planet formed at 100 K. The reason why there is more oxygen at 100 K in chemical model 1 compared to chemical model 2 are the carbon grains present in chemical model 2. Carbon takes a significant fraction of mass of the formed planet, which is taken up by oxygen bearing material in chemical model 1, resulting in a larger oxygen content in planets formed assuming chemical model 1. This then results in a larger  $\Delta$[Fe/H]/$\Delta$[O/H] fraction. Nevertheless $\Delta$[Fe/H]/$\Delta$[O/H] at 200 K and 100 K are separated by a factor of $\sim 2$ for both chemical models.

\subsection{CO ice line}

\begin{figure*}
 \centering
 \includegraphics[scale=0.7]{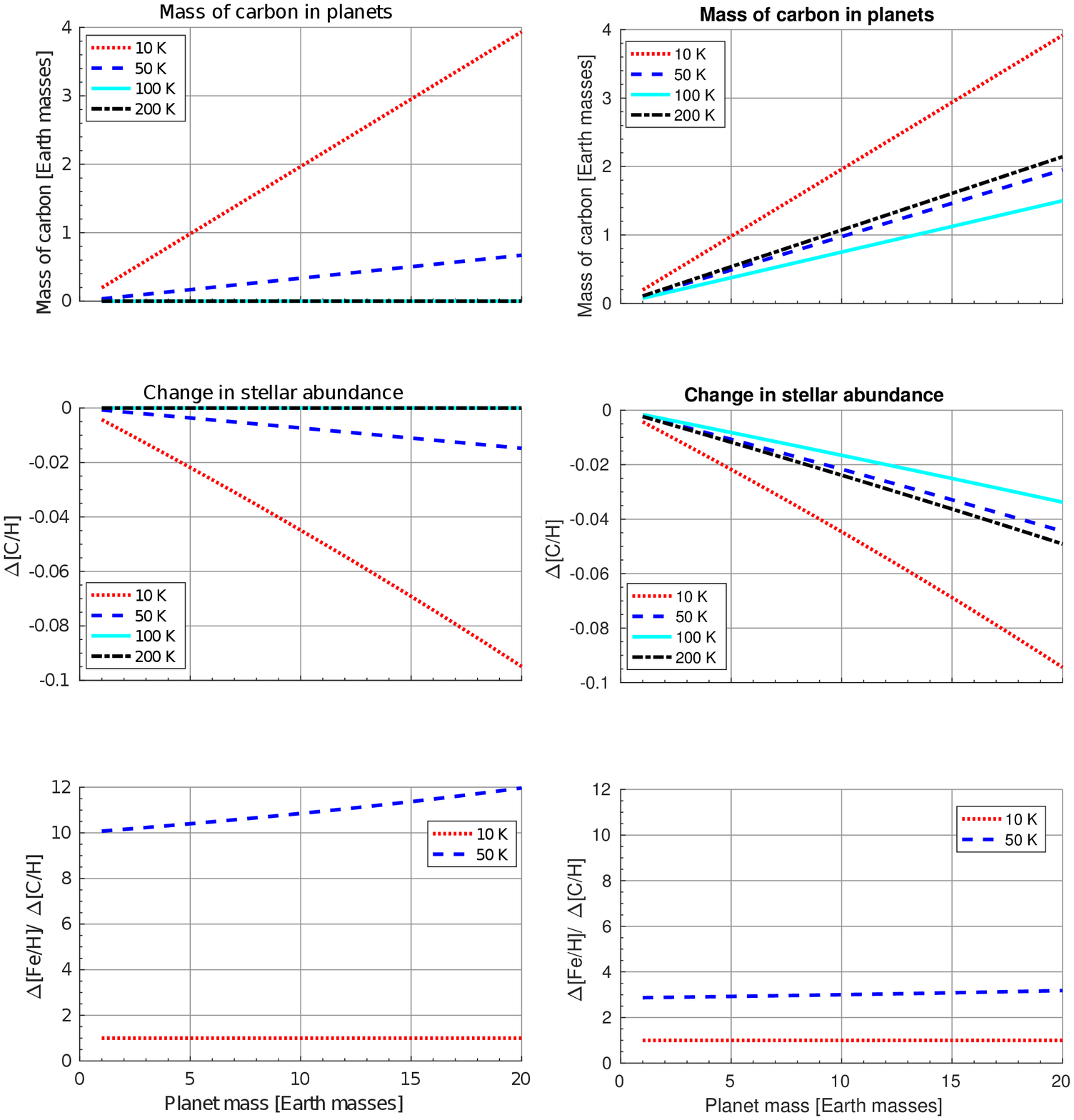} 
 \caption{Mass of carbon in the planets as function of planetary mass (top). Carbon abundance difference in the host star $\Delta$[C/H] as a function of planetary mass (middle). Ratio of the iron abundance difference $\Delta$[Fe/H] to the carbon abundance difference $\Delta$[C/H] as function of planetary mass (bottom). All plots are for planets forming at different disc temperatures and for both chemical models, where chemical model 1 is on the left and chemical model 2 on the right.
   \label{fig:Cchem}
   }
\end{figure*}

To distinguish if an ice giant has formed inside or outside the CO ice line, the change of the carbon abundance of the host star should be observed. Super Earths and ice giants forming at $r>r_{\rm CO}$ contain all volatiles in solid form and thus accrete a significant amount of CO (see table~\ref{tab:species2}), which enhances their carbon fraction significantly. In Fig.~\ref{fig:Cchem} we show the mass of carbon in the planets (top), the stellar abundance difference of carbon (middle) and the ratio of the iron abundance difference $\Delta$[Fe/H] to the carbon abundance difference $\Delta$[C/H] (bottom) for both chemical models.

Due to the unknown original disc mass and stellar convective zone in observations, we can not use the pure $\Delta$[C/H] in the stellar atmosphere to distinguish between a formation inside or outside of the CO ice line. However, the ratio of the  iron abundance difference $\Delta$[Fe/H] to the carbon abundance difference $\Delta$[C/H] is about a factor of 2 (or even more) larger for planets forming at $r<r_{\rm CO}$ ($T_{\rm disc} = 50$ K model) compared to planets forming at $r>r_{\rm CO}$ ($T_{\rm disc} = 10$ K model), independent of the chemical model, allowing to distinguish the formation location of planets.

\subsection{Migration and multiplicity}

Small mass planets in protoplanetary discs interact gravitationally with the disc and exchange angular momentum with the disc, which causes the planet to move within the disc, which is called planet migration \citep{1997Icar..126..261W}. Taking just the contribution from the Lindblad torque and the corotation torque without thermal effects into account, the time-scale for planetary migration is much shorter than the disc's lifetime, indicating that all planets would migrate towards the inner edge of the disc before disc dissipation \citep{2002ApJ...565.1257T}. Migrating planets could thus accrete material from the protoplanetary discs, which would in principle allow a planet formed at $r>r_{\rm ice}$ to migrate inwards and accrete non-icy material, especially planetesimals, which would dilute the here proposed mechanism. However, N-body simulations by \citet{1999Icar..139..350T} have shown that migrating protoplanets of a few Earth masses rather scatter planetesimals than accrete them. In the pebble accretion scenario \citep{2010A&A...520A..43O, 2010MNRAS.404..475J, 2012A&A...544A..32L}, the accretion of solid material is so fast that the planet only starts to migrate inwards significantly when it has reached its pebble isolation mass and stopped accreting solids \citep{2014A&A...572A..35L, 2015A&A...582A.112B}. We thus believe, that a planet crossing an ice line after reaching pebble isolation mass does not accrete a significant amount of solids any more, allowing a distinction through its chemical composition.

Additionally, planet migration can be directed outwards, if the radial gradients in temperature and entropy are very steep \citep{2008ApJ...672.1054B, 2011MNRAS.410..293P}, which also depends on the underlying disc model. Around the water ice line, the disc structure changes due to a transition in opacity and allows outward migration of low mass planets in many disc models \citep{2013A&A...549A.124B, 2014A&A...564A.135B, 2015A&A...575A..28B, 2015arXiv150303352B}. Importantly, planets are trapped at $r>r_{\rm ice}$, separating the regions of planet formation at the water ice line, and thus separating the accretion reservoirs of forming planets, which is represented in their chemical composition. As the disc evolves in time, the region of outward migration moves closer to the central star and allows only smaller planets to be trapped after a few Myr \citep{2015A&A...582A.112B, 2016A&A...590A.101B}. Planets of a few Earth masses can then migrate towards the inner system to orbits typical of super-Earths ($<$100 days). Recent simulations have also shown that even in disc models where outward migration does not exist that a local accretion before the planet crosses the ice line is possible \citep{2015A&A...582A.112B}.

On the other hand, recent N-body simulations of migrating planetary embryos that match the orbital configuration of Kepler planetary systems have shown that migrating planets pile up in resonant chains at the inner edge, where instabilities (and thus also collisions between the bodies) break these chains \citep{2017MNRAS.470.1750I}. This implies that the composition of super-Earths could indeed be a mixture of material from inside and outside the water ice line. We thus expand our model to account for the mixture of material from inside and outside the water ice line. The measurements of $\Delta$[Fe/H]/$\Delta$[O/H] or $\Delta$[Fe/H]/$\Delta$[C/H] can thus help to constrain where super Earths formed and where they preferably accreted most of their material.

\begin{figure}
 \centering
 \includegraphics[scale=0.7]{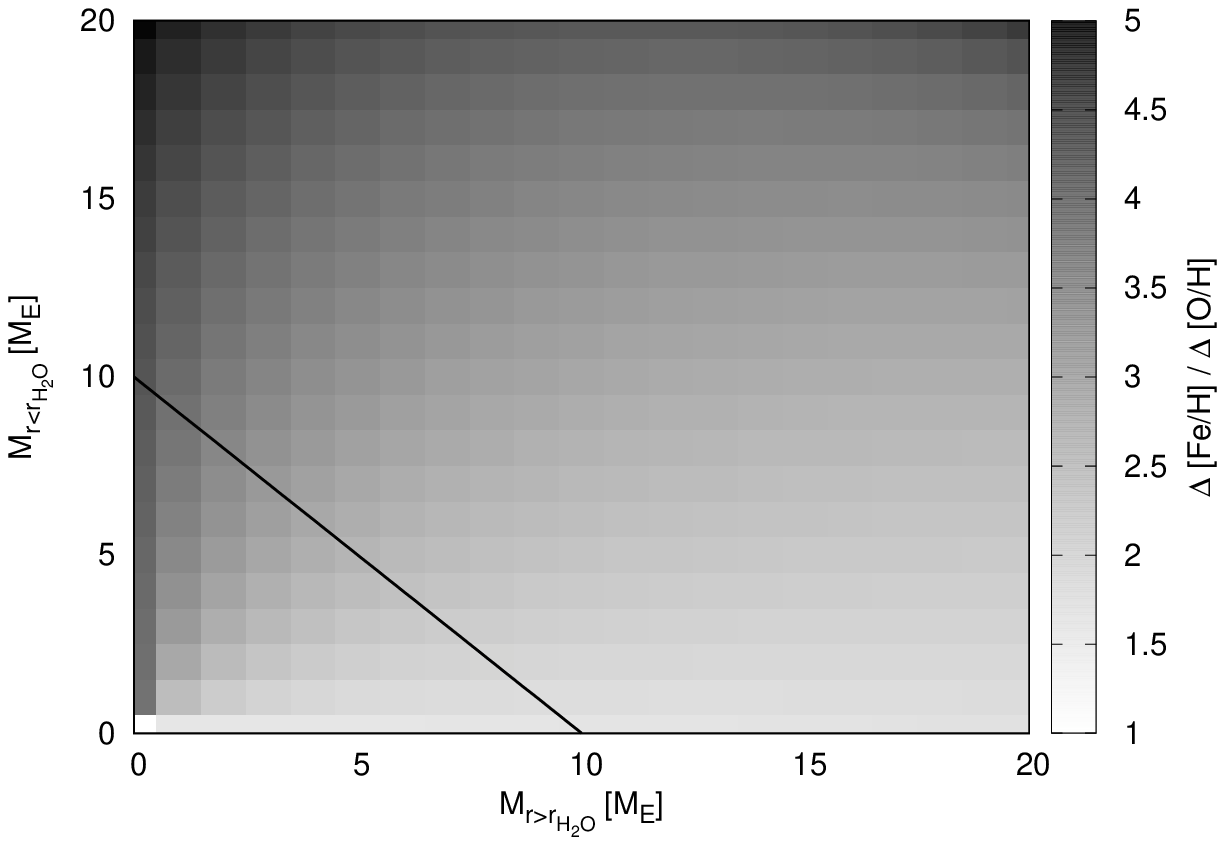} 
 \includegraphics[scale=0.7]{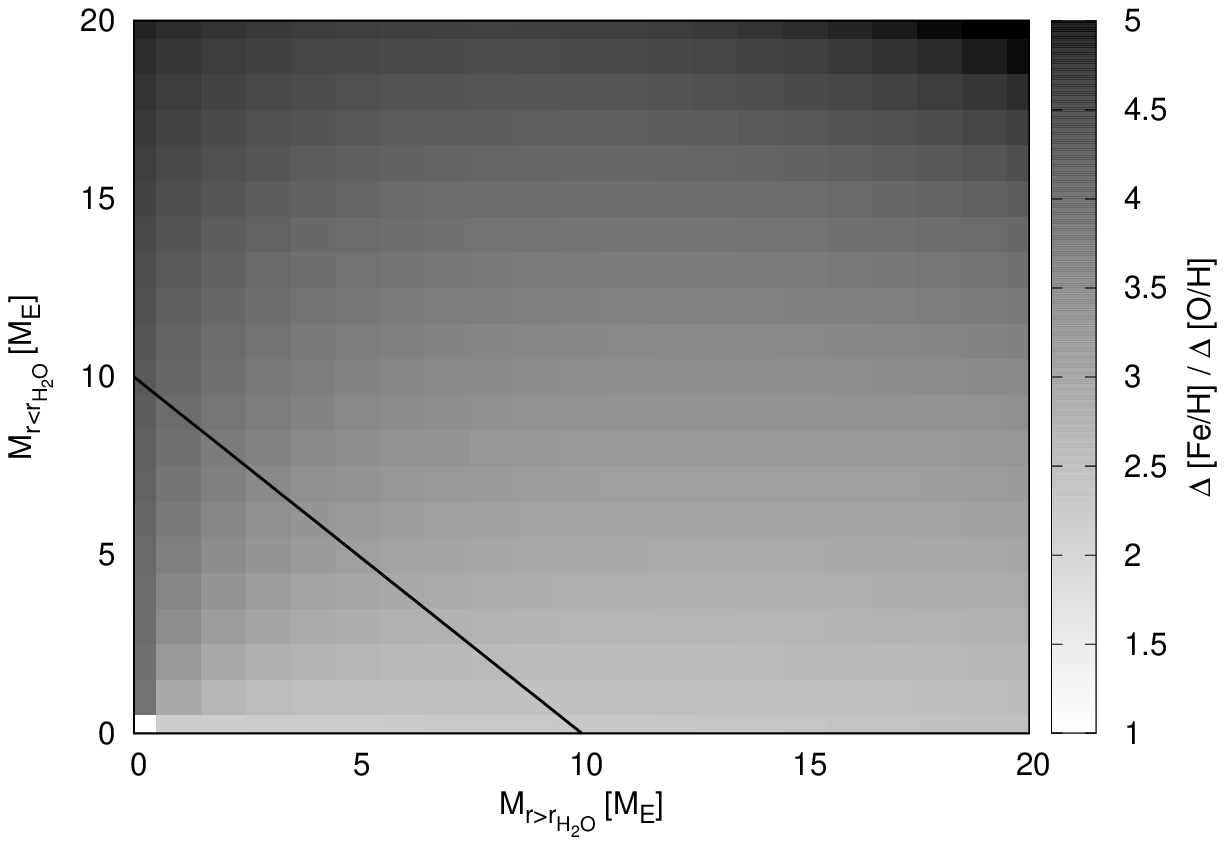}  
 \caption{Ratio of the iron abundance difference $\Delta$[Fe/H] to the oxygen abundance difference $\Delta$[O/H] as function of planetary mass for planets formed outside ($r>r_{\rm H_2O}$) and inside ($r<r_{\rm H_2O}$) the water ice line. The top plot applies to chemical model 1 and the bottom plot to chemical model 2. We note that the values for $M_{\rm r>r_{\rm H_2O}}=0$ follow the $\Delta$[Fe/H]/$\Delta$[O/H] value shown in Fig.~\ref{fig:Ochem} for the planet formed at 200 K, while the values for $M_{\rm r<r_{\rm H_2O}}=0$ follow the $\Delta$[Fe/H]/$\Delta$[O/H] shown in Fig.~\ref{fig:Ochem} for the planet formed at 100 K. The black diagonal line indicates the possible chemical abundance difference of the host star caused by the formation of planet(s) containing 10 Earth masses of solids.
   \label{fig:mixH2O}
   }
\end{figure}

In Fig.~\ref{fig:mixH2O} we show the $\Delta$[Fe/H]/$\Delta$[O/H] values for planets, which have formed partly inside or outside the water ice line. We use here a solar metallicity, [Fe/H]=0. The axes in this figure give the amount of solid material of the planet that originated from inside ($r<r_{\rm H_2O}$) or outside ($r>r_{\rm H_2O}$) the water ice line. The solid black line indicates the possible chemical abundance difference of the host star caused by a 10 Earth mass planet or by 2 or more planets with a total mass of 10 Earth masses. We want to emphasise here that the total number of planets in the system does not influence our results, as long as the total masses of all the planets in the systems are known.

If the planet is formed either completely inside or outside the water ice line, the trends in Fig.~\ref{fig:mixH2O} follow directly the values obtained in Fig.~\ref{fig:Ochem}. Only if planetary material originates from both, inside and outside the water ice line, does a deviation from the previous calculations occur. Fig.~\ref{fig:mixH2O} thus gives an overview of possible outcomes of stellar abundance difference in the stellar atmosphere related to the formation location of the planet. Clearly any observational constraint will be useful to constrain planet formation simulations, as they will either indicate that super Earths form outside or inside the water ice line or from both reservoirs.

Fig.~\ref{fig:mixH2O} shows also the difference between our two chemical models, which are very weak, as already indicated in Fig.~\ref{fig:Ochem}. We show a similar study for planets formed around the CO ice line in appendix~\ref{ap:COline}.

\section{Caveats and discussion}
\label{sec:discuss}

Additionally to planet formation, other mechanisms could influence the chemical composition of the planetary host star during the accretion of the protoplanetary disc. This could apply e.g. to the formation of planetesimal belts like the asteroid belt or through an incomplete accretion of the disc's material. We discuss in the following the influence of the protoplanetary disc mass, the stellar convective zone, planet migration, multiplicity, giant planet companions and the limitations through detection.

\subsection{Influence of the disc mass and the stellar convective zone}
\label{subsec:convect}

Planets growing in discs take material away from the disc, while the rest of the disc is accreted onto the central star. The relative fraction of the material a planet takes away from the disc to the mass of the disc itself changes with disc mass ($M_{\rm P} / M_{\rm disc}$). A planet of the same mass takes comparatively less material away from a more massive disc than from a less massive disc. However, as the remainders of the disc are accreted onto the central star, the disc mass does not influence the relative change of abundances $\Delta$[X/H] of the host star, because the relative change of the stellar abundance is determined by the mass of the convective zone (eq.~\ref{eq:deltaXH}).

A smaller convective zone results in a larger change of relative abundance $\Delta$[X/H] for a planet formed with a given mass compared to a larger stellar convective zone. As the mass of the convective zone at the beginning of the disc accretion is not known \citep{2017A&A...599A..49K}, just measuring $\Delta$[X/H] in the stellar atmosphere for a planet of a given mass is degenerate. But the ratios $\Delta$[Fe/H]/$\Delta$[O/H] and $\Delta$[Fe/H]/$\Delta$[C/H] stay constant, independently of the mass of the convective zone. However, if the convective zone is very large, the relative change of abundance $\Delta$[X/H] might be below the detection limit of $\sim 0.01$ dex with current telescopes. Nevertheless, observations of binary star systems hosting giant planets have shown that abundance differences in the order of $\sim 0.01$ dex exist and can be related to giant planet formation \citep{2014ApJ...790L..25T, 2015ApJ...808...13R, 2016ApJ...819...19T}.

\subsection{Small gaseous atmospheres}

The ice giants in our own solar system have core masses exceeding 10-15 Earth masses and small gaseous envelopes (2-5 Earth masses) around them. From the standard core accretion scenario it is also unlikely that planets in that mass regime have no gaseous envelope around them. The heavy element enrichment of Neptune and Uranus' atmospheres was measured to be roughly 20-50 times the solar enrichment \citep{1991uran.book..147F}. We thus want to discuss here the influence that a small gaseous atmosphere has on our abundance predictions. We will illustrate this with an example of a planet of 20 Earth masses with a 15 Earth mass core and 5 Earth mass gaseous envelope. This envelope is either enriched by a factor of 20 or 50 compared to solar or follows the solar abundance pattern directly. We show the results of these calculations in table~\ref{tab:envelope}.

\begin{table*}
\centerline{\begin{tabular}{c|c|c|c|c}
\hline
 & T=200 K & T=100 K & T=50 K & T=10 K \\ 
 & $\Delta$[Fe/H]/$\Delta$[O/H] & $\Delta$[Fe/H]/$\Delta$[O/H] & $\Delta$[Fe/H]/$\Delta$[C/H] & $\Delta$[Fe/H]/$\Delta$[C/H] \\ \hline \hline
20 M$_{\rm E,s}$ & 4.973 & 1.689 & 11.969 & 1.000 \\
15 M$_{\rm E,s}$+ 5 M$_{\rm E,g,1}$ & 4.508 & 1.644 & 10.962 & 1.000 \\
15 M$_{\rm E,s}$+ 5 M$_{\rm E,g,20}$ & 3.940 & 1.571 & 6.501 & 1.000 \\
15 M$_{\rm E,s}$+ 5 M$_{\rm E,g,50}$ & 3.275 & 1.468 & 3.921 & 1.000 \\
 \hline 
\end{tabular}}
\caption{Abundance ratios of planets formed interior or exterior of the water and CO ice line using chemical model 1, where the planet is either completely formed from solid material, or contains an envelope of 5 Earth masses. The envelope in itself contains heavy elements either in solar abundance or is enriched compared to solar by a factor of 20 or 50. The enrichment factor is donated by the subscript 1, 20 or 50 in the gas component. There are no changes for the planet forming at T=10 K, because all heavy element species considered in our model are in solid form and can thus not be accreted with the gas.}
\label{tab:envelope}
\end{table*}

In table~\ref{tab:envelope} we show the $\Delta$[Fe/H]/$\Delta$[O/H] and $\Delta$[Fe/H]/$\Delta$[C/H] values for a 20 Earth mass planet formed at different disc temperatures with different envelope and envelope enrichment values. The mass of the respective elements in the planetary envelope are calculated following eq.~\ref{eq:deltaXH}, because we assume that the gas accreted by the planet is below hydrogen dissociation temperature. We additionally assume that the planetary envelope was accreted at the same temperature as the solid material, meaning that the molecular species that is not in solid form at planet formation temperature is in gaseous form and will be accreted into the envelope.

Generally, a planet of the same mass, but with a gaseous envelope results in a reduced $\Delta$[Fe/H]/$\Delta$[X/H] value compared to a planet without a planetary atmosphere for the investigated formation temperatures. This is caused by an interplay between the lower iron abundance of the planet with atmosphere as well as a change of the oxygen or carbon abundance, where the amount of oxygen and carbon is in most cases also reduced, except for the large atmospheric enrichment factors. Nevertheless, the reduction of the iron abundance dominates the trend of the reduction of the $\Delta$[Fe/H]/$\Delta$[X/H] values. Additionally, the $\Delta$[Fe/H]/$\Delta$[X/H] values decrease if the enrichment of the planetary atmosphere increases. This is caused by the increased oxygen and carbon abundance in the planetary atmosphere.

In general, the trend of distinguishing the formation location of planets formed inside/outside the water ice line remain intact, with at least a factor of 2 between the $\Delta$[Fe/H]/$\Delta$[O/H] values for the formed planets. The planet forming at 50 K, on the other hand shows very large changes of the $\Delta$[Fe/H]/$\Delta$[C/H] values for increasing envelope enrichments. This is caused by the initial very low carbon abundance in solid form in the planet, where an enriched planetary atmosphere can suddenly contain more carbon than the core itself. For the planet forming at T=10 K, we observe no change in the $\Delta$[Fe/H]/$\Delta$[C/H] values with planetary atmosphere, because the atmosphere in itself does not contain any heavy elements as all heavy elements are frozen out into solids in our model. If the atmosphere would nevertheless contain Carbon, the $\Delta$[Fe/H]/$\Delta$[C/H] ratio would drop below 1, allowing an easier distinction where the planet formed.

Only when the planetary atmosphere becomes similar in mass to the planetary core mass, might the predicted trends not be observable any more. However, if the atmosphere is nearly as massive as the core mass, the planet is close to transitioning into the runaway gas accretion phase and becoming a gas giant \citep{1996Icar..124...62P}.

The trends here described indicate that even if the planets have a small planetary atmosphere, the trends in $\Delta$[Fe/H]/$\Delta$[X/H] remain valid and can be used to distinguish if the planets formed interior or exterior to the water or CO ice line.

\subsection{Multiplicity and  giant planet companions}
\label{subsec:multi}

Our simulations predict that the abundance difference of heavy elements in host stars compared to their binary partners due to planet formation in protoplanetary discs is observable, if the formed planets have reached a mass of about 10 Earth masses. This mass is typically a bit too massive for a standard super-Earth, however super-Earth are mostly in systems of multiple planets \citep{2013ApJ...766...81F}. In case there are several super-Earths in one system, the distinction between the formation inside or outside of the water ice line works in the same way, however, then all of the observed super-Earths have to form either inside or outside the water ice line for our predictions to follow Fig.~\ref{fig:Ochem} or Fig.~\ref{fig:Cchem}. In case some super-Earths form inside and some super-Earths form outside of the water ice line or the super Earth in itself crossed an ice line during its accretion, the predicted value of $\Delta$[Fe/H]/$\Delta$[O/H] discussed in section~\ref{sec:results} will be diluted. However, the change of abundance is then in between the predicted abundance changes (Fig.~\ref{fig:mixH2O}), which can be observed, if the measurement errors are small enough and the total mass of the planets is large enough. This would then imply that in the same system, super-Earths can form inside and outside of the water ice line, which is also a valuable information for planet formation theories.

If the detection of planets is incomplete, we will miss valuable information that can influence our predictions. Several missing or more massive missed companions dilute our predictions more significantly, which is why the observed system has to be observationally as complete as possible. On the other hand, analysis of the Kepler samples have shown that planets within the same system have mostly the same planetary radius \citep{2018AJ....155...48W} and thus presumably mass, allowing a prediction from the observed abundance difference in $\Delta$[Fe/H]/$\Delta$[O/H]. For example, if $\Delta$[Fe/H]/$\Delta$[O/H]=3.0 is measured, but only one planet with 6 Earth masses is found, Fig.~\ref{fig:mixH2O} indicates that another planet of 4 Earth masses is in the system, when using chemical model 1. However, we would then be unable to determine if one planet formed completely inside the water ice line and one planet completely outside the water ice line or if both planets contain a mixture of material. Nevertheless, this would be an indication that the building blocks of super Earths originate from both inside and outside the water ice line.

In contrast to super-Earths, which are mostly rocky, giant planets also accrete significant amounts of gas from the protoplanetary disc, which depletes the volatile components inside the disc, influencing in the end the measured abundances of the host star. Therefore, systems hosting giant planets can easily influence our predictions. However, systems of super Earth mostly do not have giant planet companions, also because of the simple fact that giant planets are very rare \citep{2011arXiv1109.2497M}, while super Earth systems are very common \citep{2013ApJ...766...81F}. Additionally, searching in metal poor systems might evade the potential contamination of our results due to (unseen) giant planets, because metal poor systems host far fewer giant planets than metal rich systems \citep{2005ApJ...622.1102F, 2010PASP..122..905J}, while there is no such correlation for super Earths. This reduces the probability to have giant planets in the system even further, allowing a better conclusion from the observational data.

\subsection{Detection limits}

Our method requires the detection of planets with the RV method, because the planetary masses are needed. However, planet detections with the RV method, only give minimum masses of such planets, which introduces a bias in our model, because a 10 ${\rm M}_{\rm E}$ formed at $r>r_{\rm ice}$ gives $\Delta$[O/H]=-0.05, while a 16 ${\rm M}_{\rm E}$ formed at $r<r_{\rm ice}$ gives the same change of oxygen abundance in the host star (chemical model 1 in Fig.~\ref{fig:Ochem}). Additional constraints on the planet mass compared to just RV data might thus be needed, for example through transit observations.

The observed binary star needs to be spectroscopically distinct and the binary separation should be wide enough ($>$100-200 AU), so that the protoplanetary discs around the individual stars in the binary system do not interchange mass. 

Additionally, the binary stars should be both about solar mass, because only then the convective envelope of the star is about a few percent, making the changes of the stellar abundance due to the volatile depleted material from the disc due to planet formation observable. Therefore only stars with $0.8 {\rm M}_\odot < M_\star < 1.3 {\rm M}_\odot$ can be used for these observations.

It is important to note that {\it both} stars in the stellar binary should be spectral twins, because otherwise it is observationally difficult to resolve abundance difference of the 0.01 dex level due to significantly different stellar temperatures and surface gravities. In addition, the photospheric diffusion may change surface abundances differently for stars of different spectral types as they age \citep{2017ApJ...840...99D}. In addition, the theoretical assumptions of our model (e.g. similar evolution of the convective zone) might not be valid if the stars in the binary are not twins. Therefore to have best results for our prediction it is important to observe twin binaries.

%- Section 5.3: This may be a good place to clarify whether these results apply to all FGK stars in wide binaries or only to spectral twins in binaries. All of the previously studied binaries with detected abundance differences have been twins, because it is observationally difficult to resolve 0.01 dex level changes between stars with significantly different temperatures/gravities. Additionally, photospheric diffusion may change surface abundances differently for stars of different spectral types as they age (e.g. Dotter et al 2017ApJ...840...99D). For these reasons, I suspect that twin binaries will be required for these observations. I also wonder whether the theoretical assumptions (that the size of the stellar convective zone at the time of disc accretion is comparable for both stars, that the fraction of gas lost to stellar winds is similarly negligible for both, etc) remain valid for non-twin stars.

Field solar twins (i.e. stars which are indistinguishable with the Sun in terms of their effective temperature, surface gravity and metallicity) are not ideal targets for our predictions, because Galactic chemical evolution \citep{2014A&A...564L..15A}, slightly different birth places and thus different initial chemical composition of the stars, as well as different stellar ages \citep{2015A&A...579A..52N} can influence our results. In contrast, binary stars form from the same parent molecular cloud and thus have the same chemical composition, making a prediction regarding the formation location of planets via stellar abundance differences more reliable.

There are several test cases for our predictions, HD 20781/20782 (hosting a Jupiter planet around component A, and two Neptune mass planets around component B), Kepler-449 (two objects with 2-3 R$_{\rm E}$ around component A), EPIC201629650 (one object with 2 R$_{\rm E}$ around component A), EPIC201384232 (one object with 2.5 R$_{\rm E}$ around component A), EPIC201403446 (one object with 2.1 R$_{\rm E}$ around component A). We note here that HD 20781/20782 might not be the most ideal test case for our predictions, even though high resolution spectroscopic data of the host star exists \citep{2014ApJ...787...98M}, because the error bars are large ($\sim 0.07$ dex) and both stars have a difference in effective temperature of $T_{\rm eff} \sim 500$ K. Additionally, the Jupiter type planet additionally accretes gas, which is not taken into account in our model.
 
High precision stellar spectroscopy for the other binary stars does not exist up to date, so these planetary systems are an ideal test for our predictions, because they additionally do not show evidence of giant planets. As the masses of these planets are still unknown, they can be either inferred roughly from mass-radius relationships \citep{2016ApJ...825...19W} or need to be determined by follow up RV measurements as well for our predictions to work best.

The observed abundance differences in binary stars could be due to planet formation. Although the signature could be diluted or mixed with other factors, it is still likely to observe the predicted abundance ratios. Gaia DR2 will release a large number of co-moving solar-type binaries which will be ideal targets to test the proposed scenario and the predicted difference of abundance ratios in addition to the before mentioned systems.

\section{Summary}
\label{sec:summary}

The ways super-Earths are born is still a mystery and their formation depends on their growth and migration pathways. Observational constraints additionally give information about the period ratios of super-Earths, constraining models of their growth and migration history. Here, we suggest to use the observation of stellar abundances in binary star systems to disentangle the birth environment of super-Earths and also ice giants. These observations will be a very important piece of the puzzle to explain where and how super-Earths and ice giants are formed.

Forming stars are surrounded by protoplanetary discs, which accrete onto the central star. As planets form in the disc, they take solid material out from the disc, meaning that the leftover material accreted onto the central star is depleted in volatiles and refractories. This results in a decreased stellar abundance due to the material incorporated into planets compared to the non-planet hosting stars. In binary star systems, where one stellar companion hosts planets, while the other one is without planets, a difference in stellar abundance is thus observable. This difference in stellar abundance, however, depends strongly on the formation location of the formed planet. In our work we have calculated this abundance difference for super-Earths and ice giants\footnote{We assume that both planetary types consists mostly of solids, but see section 5, where we also discuss the influence of a small gaseous atmosphere.} for two different chemical models.

Planets forming at $r<r_{\rm H_2O}$, so in the hot parts of the protoplanetary discs, harbour significantly less oxygen (because they do not accrete water) compared to planets forming at $r>r_{\rm H_2O}$. This leads to a difference in the oxygen abundance of the host star, because the material accreted from the protoplanetary disc is depleted in oxygen for the planet forming at $r<r_{\rm H_2O}$ compared to the planet forming at $r>r_{\rm H_2O}$. For a planet of the same mass, the difference in the oxygen abundance is larger for the planet formed at $r>r_{\rm ice}$.

Additionally, the host stars show a difference in the abundance of iron, sulphur, magnesium and silicon if planets formed inside or outside the water ice line. This is caused by the fact that planets formed at $r<r_{\rm H_2O}$ accrete no water, but if they have the same mass, they consist to a larger fraction of iron, sulphur, magnesium and silicon. These differences are also observable and in combination with the oxygen abundances these measurements can constrain if a planet formed inside or outside the water ice line (see Fig.~\ref{fig:Ochem}).

A similar process applies to distinguish if ice giants formed inside or outside the CO snow line. Here, however, carbon becomes of interest, because planets forming at $r>r_{\rm CO}$ contain a lot of CO ices, while planets forming $r<r_{\rm CO}$ contain no CO ice. This leads to a much lower carbon abundance difference in the atmosphere of the host star compared to a planet forming at $r>r_{\rm CO}$. These changes are again observable and also in combination with the abundance differences of iron, sulphur, magnesium and silicon allow to constrain if an ice giant formed inside or outside the CO snow line.

The relative trends of $\Delta$[Fe/H]/$\Delta$[O/H] to determine if a planet formed inside or outside the water ice line and $\Delta$[Fe/H]/$\Delta$[C/H] to determine if a planet formed inside or outside the CO ice line are independent of the original mass of the protoplanetary disc where the planet(s) formed and of the mass of the convective zone of the central star. For small convective zones, the relative abundance difference of an element $\Delta$[X/H] becomes larger and is thus easier to measure, while for large convective zones $\Delta$[X/H] might become lower than the detection limit of $\sim 0.01$ dex, but stellar spectroscopy analysis of binary star systems hosting planets already show small abundance differences \citep{2014ApJ...790L..25T, 2015ApJ...808...13R, 2016ApJ...819...19T}. Our model also accounts for mixing effects, if planets formed partly inside or outside an ice line (Fig.~\ref{fig:mixH2O}) allowing to constrain the formation location of super Earths. Additionally, our model is applicable to multi planet systems, if the total mass of the planets is known.

So far, observations have only considered binary star systems with giant planets to constrain their formation pathway, where already some conclusions about their formation location can be made \citep{2014ApJ...790L..25T, 2015ApJ...808...13R, 2016ApJ...819...19T}. Our method can thus allow to extend this search to super-Earths and ice giants, where future observation of binary star systems hosting super-Earths or ice giants will give valuable information to constrain the theory of the formation of those planets.

\section*{Acknowledgements}

B.B., thanks the European Research Council (ERC Starting Grant 757448-PAMDORA) and the Knut and Alice Wallenberg Foundation (grant 2012.0150) for their financial support. A.J.\, thanks the Knut and Alice Wallenberg Foundation (grants 2012.0150), the Swedish Research Council (grant 2014-5775) and the European Research Council (ERC Consolidator Grant 724687-PLANETESYS) for their financial support. F.L. was supported by the Swedish Research Council (grant 2012-2254). We additionally thank the referee for his/her comments that helped us to improve our manuscript.

\appendix
\section{Multiplicity at the CO ice line}
\label{ap:COline}

\begin{figure}
 \centering
 \includegraphics[scale=0.7]{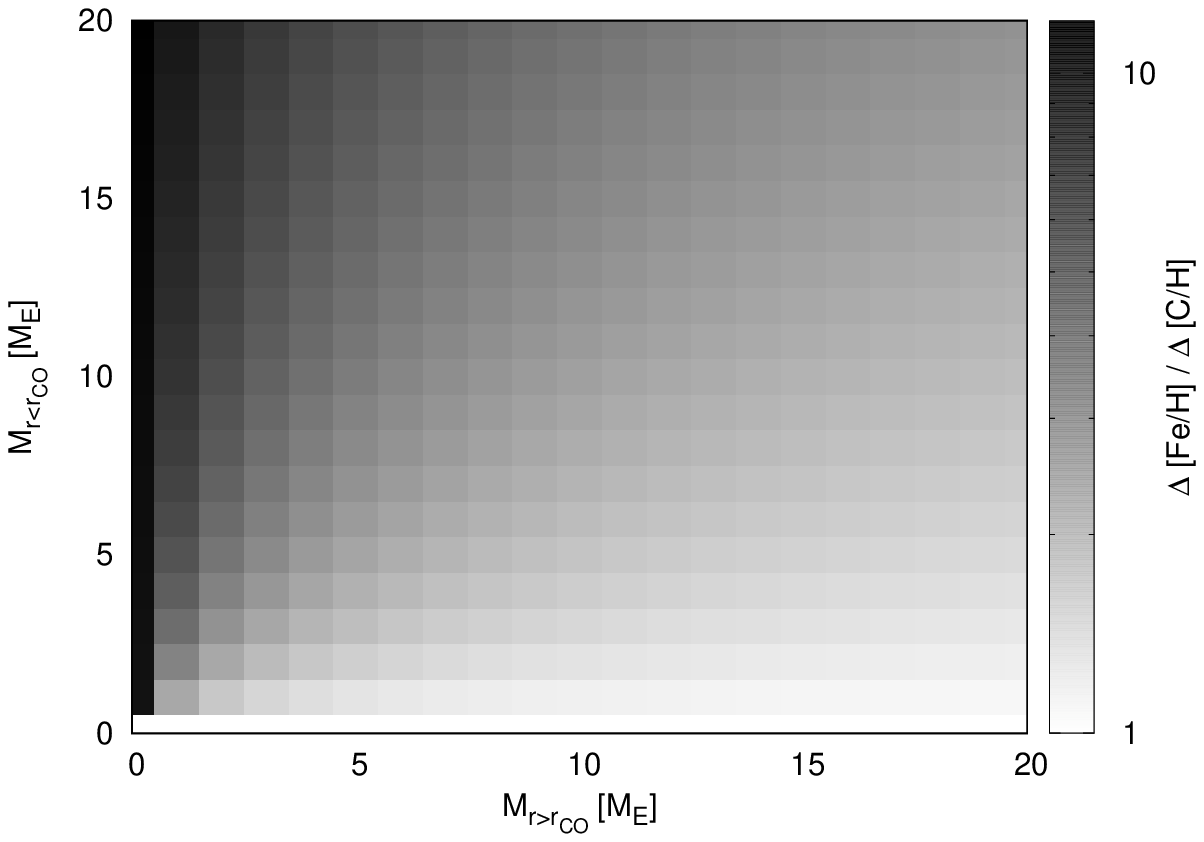} 
 \includegraphics[scale=0.7]{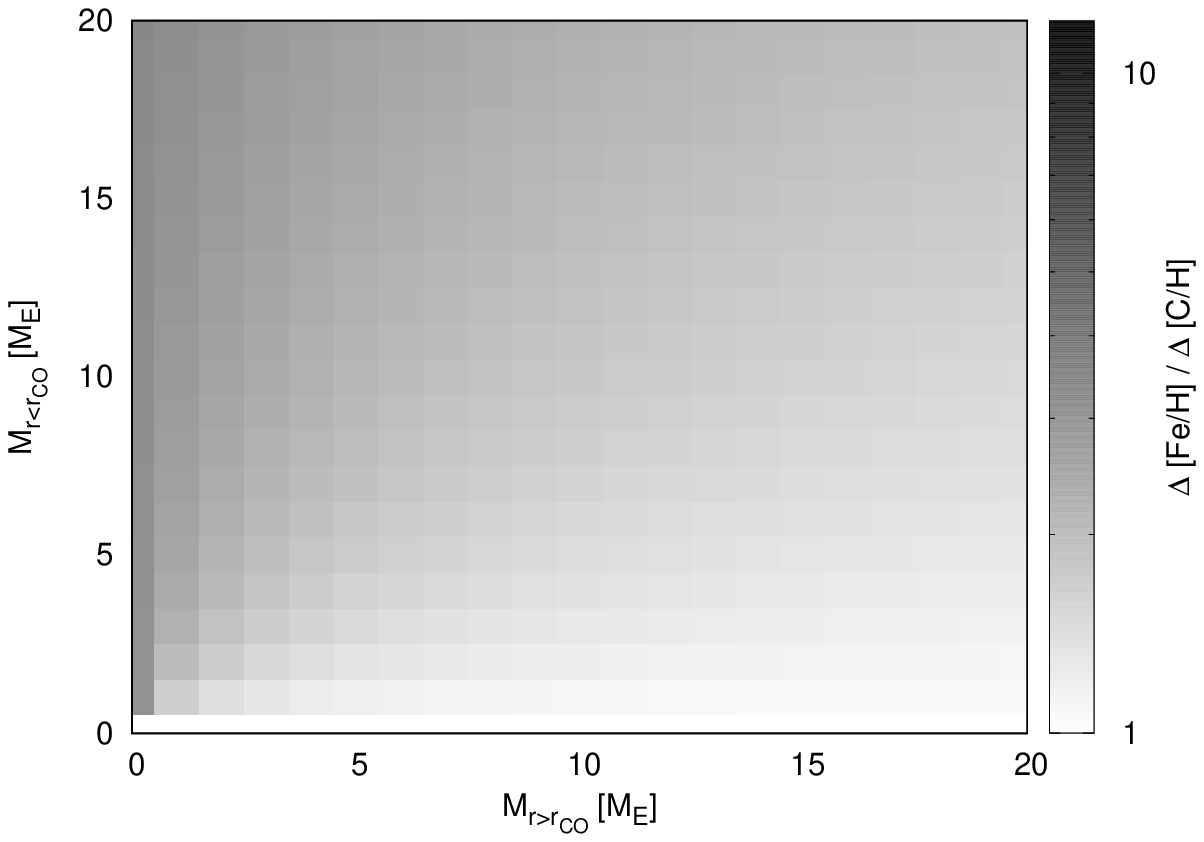}  
 \caption{Ratio of the iron abundance difference $\Delta$[Fe/H] to the oxygen abundance difference $\Delta$[C/H] as function of planetary mass for planets formed outside ($r>r_{\rm CO}$) and inside ($r<r_{\rm CO}$) the water ice line. The top plot applies to chemical model 1 and the bottom plot to chemical model 2. We note that the values for $M_{\rm r>r_{\rm CO}}=0$ follow the $\Delta$[Fe/H]/$\Delta$[C/H] value shown in Fig.~\ref{fig:Cchem} for the planet formed at 50 K, while the values for $M_{\rm r<r_{\rm CO}}=0$ follow the $\Delta$[Fe/H]/$\Delta$[C/H] shown in Fig.~\ref{fig:Cchem} for the planet formed at 10 K.
   \label{fig:mixCO}
   }
\end{figure}

We show in Fig.~\ref{fig:mixCO} the $\Delta$[Fe/H]/$\Delta$[C/H] values for planets, which can have formed partly inside or outside the CO ice line. We use a solar metallicity, [Fe/H]=0. The axes in this figure give the amount of solid material of the planet that originated from inside ($r<r_{\rm CO}$) or outside ($r>r_{\rm CO}$) the CO ice line. We want to emphasise here that the total number of planets in the system does not influence our results, as long as the total masses are known. Clearly, as already indicated in Fig.~\ref{fig:Cchem}, the different chemical models result in a very different abundance difference in the host star's atmosphere. Nevertheless, both chemical models allow to distinguish if a planet formed completely inside or outside the CO ice line or if its material is a mixture.

\bibliographystyle{mnras}
\bibliography{Stellar}
\label{lastpage}
\end{document}